\documentclass[aps,prd,superscriptaddress,preprintnumbers,amsmath,amssymb,latexsym,array,enumerate,letterpaper,floatfix,nofootinbib]{revtex4-1}
\usepackage{epsfig}
\usepackage[colorlinks=true,urlcolor=brown,linkcolor=blue,citecolor=magenta]{hyperref}
\usepackage{breakurl}
\usepackage{bm,xcolor}
\usepackage{slashed}
\usepackage{cancel}
\usepackage{graphicx}
\usepackage{multirow}
\usepackage{comment}
\usepackage{epstopdf}
\usepackage{mathtools}
\usepackage{gensymb}
\usepackage{appendix}
\usepackage{url}
\usepackage[english]{babel}

\usepackage[skip=2pt plus1pt, indent=15pt]{parskip}

\makeatletter
\def\simgt{\mathrel{\lower2.5pt\vbox{\lineskip=0pt\baselineskip=0pt
           \hbox{$>$}\hbox{$\sim$}}}}
\def\simlt{\mathrel{\lower2.5pt\vbox{\lineskip=0pt\baselineskip=0pt
           \hbox{$<$}\hbox{$\sim$}}}}
\makeatother

\newcommand{\be}{\begin{equation}}
\newcommand{\ee}{\end{equation}}
\newcommand{\bea}{\begin{eqnarray}}
\newcommand{\eea}{\end{eqnarray}}
\newcommand{\beq}{\begin{eqnarray}}
\newcommand{\eeq}{\end{eqnarray}}
\newcommand{\Fig}[1]{Fig.~(\ref{#1})}
\newcommand{\Eq}[1]{Eq.~(\ref{#1})}
\newcommand{\Eqs}[2]{Eqs.~(\ref{#1}) and (\ref{#2})}
\newcommand{\Sec}[1]{Sec.~\ref{#1}}

\newcommand{\App}[1]{App.~\ref{#1}}

\newcommand{\eg}{\textit{e.g.}}
\newcommand{\ie}{\textit{i.e.}}

\newcommand{\keV}{\mathrm{keV}}
\newcommand{\MeV}{\mathrm{MeV}}
\newcommand{\GeV}{\mathrm{GeV}}
\newcommand{\TeV}{\mathrm{TeV}}
\newcommand{\gr}{\mathrm{g}}

\newcommand{\kpc}{\mathrm{kpc}}
\newcommand{\Mpc}{\mathrm{Mpc}}

\newcommand{\cm}{\mathrm{cm}}

\newcommand{\km}{\mathrm{km}}

\newcommand{\Gyr}{\mathrm{Gyr}}
\newcommand{\yr}{\mathrm{yr}}

\newcommand{\SU}[1]{\mathrm{SU}(#1)}
\newcommand{\Ua}{\mathrm{U}(1)}

\newcommand{\dec}{\mathrm{dec}}
\newcommand{\ann}{\mathrm{ann}}
\newcommand{\mfp}{\mathrm{mfp}}

\newcommand{\DM}{\mathrm{DM}}
\newcommand{\pDM}{\mathrm{pDM}}
\newcommand{\trg}{\mathrm{trg}}
\newcommand{\med}{\mathrm{med}}

\newcommand{\abs}[1]{\ensuremath{\bl\vert{#1}\br\vert}}

\newcommand{\anti}[1]{\ensuremath{\overline{#1}}}

\newcommand{\bl}{\left}
\newcommand{\br}{\right}

\newcommand{\dd}{{\mathrm{d}}}

\newcommand{\BH}{\textrm{BH}}
\newcommand{\PBH}{\textrm{PBH}}

\def\lsim{\mathrel{\rlap{\lower4pt\hbox{\hskip1pt$\sim$}}
     \raise1pt\hbox{$<$}}}         
\def\gsim{\mathrel{\rlap{\lower4pt\hbox{\hskip1pt$\sim$}}
     \raise1pt\hbox{$>$}}}         

\newcommand{\ignore}[1]{}

\begin{document}

\preprint{
\begin{minipage}{5cm}
\begin{flushright}
UMD-PP-024-09 \\
FERMILAB-PUB-24-0548-T-V\\
MI-HET-838
 \end{flushright}
\end{minipage}
}

\title{Light in the Shadows: Primordial Black Holes Making Dark Matter Shine}

\author{Kaustubh~Agashe}
\email{kagashe@umd.edu}
\affiliation{Maryland Center for Fundamental Physics, Department of Physics, University of Maryland, College Park, MD 20742, USA}

\author{Manuel~Buen-Abad}
\email{buenabad@umd.edu}
\affiliation{Maryland Center for Fundamental Physics, Department of Physics, University of Maryland, College Park, MD 20742, USA}
\affiliation{Department of Physics and Astronomy, Johns Hopkins University, Baltimore, MD 21218, USA}
\affiliation{Dual CP Institute of High Energy Physics, C.P. 28045, Colima, M\'{e}xico}

\author{Jae~Hyeok~Chang} 
\email{jhchang@fnal.gov}
\affiliation{Maryland Center for Fundamental Physics, Department of Physics, University of Maryland, College Park, MD 20742, USA}
\affiliation{Department of Physics and Astronomy, Johns Hopkins University, Baltimore, MD 21218, USA}
\affiliation{Theory Division, Fermi National Accelerator Laboratory, Batavia, IL 60510, USA}
\affiliation{Department of Physics, University of Illinois Chicago, Chicago, IL 60607, USA}

\author{Steven~J.~Clark} 
\email{sclark@hood.edu}
\affiliation{Hood College, Frederick, MD 21701, USA}

\author{Bhaskar~Dutta} 
\email{dutta@tamu.edu}
\affiliation{Mitchell Institute for Fundamental Physics and Astronomy, Department of Physics and Astronomy, Texas A\&M University, College Station, TX 77845, USA}

\author{Yuhsin~Tsai} 
\email{ytsai3@nd.edu}
\affiliation{Department of Physics and Astronomy, University of Notre Dame, IN 46556, USA}

\author{Tao~Xu} 
\email{tao.xu@ou.edu}
\affiliation{Homer L. Dodge Department of Physics and Astronomy, University of Oklahoma, Norman, OK 73019, USA}

\begin{abstract}

We consider the possibility of indirect detection of dark sector processes by investigating a novel form of interaction between ambient dark matter (DM) and primordial black holes (PBHs).
The basic scenario we envisage is that the ambient DM is ``dormant'', \ie, it has interactions with the SM, but its potential for an associated SM signal is not realized for various reasons.
We argue that the presence of PBHs with active Hawking radiation (independent of any DM considerations) can act as a catalyst in this regard by overcoming the aforementioned bottlenecks.
The central point is that PBHs radiate all types of particles, whether in the standard model (SM) or beyond (BSM), which have a mass at or below their Hawking temperature. The emission of such radiation is ``democratic" (up to the particle spin), since it is based on a coupling of sorts of gravitational origin.
In particular, such shining of (possibly dark sector) particles onto ambient DM can then activate the latter into giving potentially observable SM signals.
We illustrate this general mechanism with two specific models.
First, we consider
asymmetric DM, which is characterized by an absence of ambient anti-DM, and consequently the absence of DM indirect detection signals.
In this case, 
PBHs can ``resurrect'' such a signal by radiating anti-DM, which then annihilates with ambient DM in order to give SM particles such as photons.
In our second example, we consider the PBH emission of dark gauge bosons which can excite ambient DM into a heavier state (which is, again, not ambient otherwise), this heavier state later decays back into DM and photons.
Finally, we demonstrate that we can obtain observable signals of these BSM models from asteroid-mass PBHs (Hawking radiating currently with $\sim \mathcal{O}(\MeV)$ temperatures) at gamma-ray experiments such as AMEGO-X.
\end{abstract}

\maketitle

\section{Introduction}

The nature of dark matter (DM) is arguably one of the largest open questions in particle astrophysics and cosmology. The overwhelming evidence for its existence, however, entirely relies on its gravitational effects. Nevertheless, the hope is that DM has other, non-gravitational interactions with standard model (SM) particles as well, thereby providing opportunities for detecting it via some associated SM signal. Several candidates presenting just such opportunities have emerged as some of the most promising explanations for DM. Undoubtedly the most widely-studied case is that of the weakly-interacting massive particle or WIMP (\cite{Fox:2019bgz} and references therein). Part of the reason behind its popularity is that upon thermal freeze-out, WIMPs acquire approximately the right relic abundance, the so-called WIMP ``miracle''. Perhaps the most significant attraction of the WIMP paradigm is that TeV-scale new particles are typically invoked in order to solve another problem, the Planck--weak hierarchy. If one of these new particles is stable, it will be a WIMP DM candidate with necessarily some couplings to the SM. Another well-respected DM candidate is the QCD axion, which is motivated primarily not as an explanation for DM, but as a solution to the strong CP problem of the SM (\cite{Hook:2018dlk} and references therein). Thus, the QCD axion DM candidate must of necessity couple to, at least, the QCD sector of the SM.

Significant efforts to discover WIMPs, axions, and many other DM candidates via signals generated by their interactions with the SM have been, and continue to be, a central part of the experimental program in particle physics. These efforts are based on the assumption that interactions between DM and the SM arise either from direct couplings or from couplings mediated by force carriers, thereby offering clear and incontrovertible evidence of new physics beyond the Standard Model (BSM). However, no definitive, non-gravitational DM signals have so far been detected.

We are thus confronted with the possibility that the SM's connection to the DM sector may be far weaker. This could be due to either very small direct DM--SM couplings or to something preventing the production of the force carriers that connect the DM to SM particles. Put succinctly, the DM could appear to be ``inactive" today, even if it interacted with SM particles in the early Universe. Consequently, traditional DM detection methods might not be successful despite the presence of DM--SM portal interactions and the sizable abundance of DM in the cosmos. For example, in the case of asymmetric DM (ADM) \cite{Spergel:1984re,Nussinov:1985xr,Gelmini:1986zz,Barr:1990ca,Barr:1991qn,Kaplan:1991ah,Gudnason:2006ug,Gudnason:2006yj,Kaplan:2009ag,Davoudiasl:2012uw,Petraki:2013wwa,Zurek:2013wia} there is no ambient anti-DM for DM--anti-DM annihilations to occur and produce SM signals, even if a large portal coupling exists. Thus, in the ADM framework, {\em in}direct detection signals are absent.

And yet, as we show in this paper, all is far from lost. This seemingly dark prospect, of a quiescent DM, may not be without its bright side. Indeed, the possibility of indirectly detecting DM in some fashion may remain viable if the DM is made of one or several particles or objects which are in turn merely a part of a broader, richer {\it dark sector} (DS). While a theoretically conservative outlook may see such {\it dark complexity} \cite{Asadi:2022njl,Bechtol:2022koa,Dvorkin:2022jyg} as uneconomical, the fact is that this dismissive attitude may be too rash. Ordinary matter, with its complex phenomena of galaxies, stars, and gas clouds comprised of atoms, nuclei, and ultimately first-generation quarks and leptons (not to mention their antiparticles, and the second and third generations), is undeniably rich in interesting physics, and yet it constitutes a mere 17\% of the total matter in the Universe (and only 5\% of the total energy density) \cite{Planck:2018vyg}. More so, there are currently tantalizing hints that DS may help explain increasingly-significant tensions in cosmological data \cite{Abdalla:2022yfr,Poulin:2023lkg,DiValentino:2021izs,Battye:2014qga,Buen-Abad:2015ova,Lesgourgues:2015wza,Enqvist:2015ara,Murgia:2016ccp,Kumar:2016zpg,Chacko:2016kgg,Poulin:2016nat,Buen-Abad:2017gxg,Buen-Abad:2018mas,Dessert:2018khu,Poulin:2018cxd,Buen-Abad:2019opj,Smith:2019ihp,Lin:2019qug,Alexander:2019rsc,Agrawal:2019lmo,Escudero:2019gvw,Berghaus:2019cls,Archidiacono:2019wdp,RoyChoudhury:2020dmd,Brinckmann:2020bcn,Das:2020xke,Heimersheim:2020aoc,Clark:2020miy,FrancoAbellan:2021sxk,Niedermann:2021vgd,Aloni:2021eaq,Schoneberg:2022grr,Joseph:2022jsf,Buen-Abad:2022kgf,Berghaus:2022cwf,Brinckmann:2022ajr,Sandner:2023ptm,Allali:2023zbi,Buen-Abad:2023uva,Schoneberg:2023rnx,Rogers:2023upm,Bagherian:2024obh}. These DS setups may resemble the SM in many ways, having dark atoms \cite{Kaplan:2009de,Kaplan:2011yj} or even a whole mirror SM sector \cite{Hodges:1993yb,Berezhiani:1995am,Mohapatra:2000rk,Berezhiani:2000gw,Foot:2003jt,Berezhiani:2005ek,An:2009vq,Lonsdale:2018xwd,Bodas:2024idn}; signals of such complexity may be gravitationally imprinted on cosmological observables such as the anisotropies of the cosmic microwave background and large-scale structure~\cite{Cyr-Racine:2012tfp,Cyr-Racine:2013fsa,Bansal:2021dfh,Bansal:2022qbi,Zu:2023rmc}. Finally, many BSM theories concerned with fundamental questions in high-energy physics give rise to more than one dark particle; Twin Higgs \cite{Chacko:2005pe,Chacko:2005un,Chacko:2005vw,Burdman:2014zta,Craig:2015pha}, N-Naturalness \cite{Arkani-Hamed:2016rle,Easa:2022vcw,Batell:2023wdb}, and more generally ``neutral naturalness'' \cite{Batell:2022pzc} are examples of two such theories.

Given the possibility of dark complexity within a DS, it is plausible that DM interacts with other DS particles more strongly than with SM particles. Thus, DS interactions may be necessary for producing observable signals. If the DS particles requisite for these DS interactions to take place are absent today, the DM would remain effectively dormant. Therefore, in order for visible indirect detection signals to be possible, we must identify the astrophysical environments capable of producing these ``{\it triggeron}'' DS particles necessary to trigger or activate DS processes such as DM annihilations or decays. Said differently, these triggerons are essential for initiating indirect detection signals. In the case of ADM, the absence of anti-DM in today's Universe results in no DM--anti-DM annihilations and, therefore, no SM signals. To generate visible signals, we need to first produce the triggeron, which in this case is the anti-DM. Alternatively, consider a scenario where a heavier DS particle, which emits SM photons upon decaying into DM, serves as the sole portal to the SM but exists in negligible abundance today. In such cases, a DS triggeron could excite ambient DM into this heavier DS state, thereby producing photon signals.

In these examples, the question of the observability of indirect detection signals revolves around identifying the circumstances under which a significant number of triggerons can be generated to activate the necessary DM interactions. For instance, Ref.~\cite{Agashe:2020luo} investigated the activation of DM annihilations in an ADM model by creating anti-DM through the annihilation of a different species of DM particle. An observable signal becomes feasible if these two types of DM particles collectively comprise a substantial portion of the total DM, and if the specific details of the DS model manifest in nature.

There is, however, a model-independent way to produce DS particles, and thus perhaps the required triggerons. Gravity, the one universal interaction for both visible and invisible particles, operates indiscriminately on the DS and the SM. This implies that under sufficiently strong gravitational fields, such as those near black holes (BHs), DS particles could be generated. Indeed, BHs emit particles via Hawking radiation~\cite{Hawking:1974rv,Hawking:1975vcx}, and are thus capable of producing \textit{any} particle provided its mass is not significantly greater than the BH temperature. Such a generic production channel has been used to probe light decaying DS particles~\cite{Agashe:2022phd}. Given this context, we point out that sufficiently light triggerons {\it will} be emitted by BHs through Hawking radiation, potentially triggering interactions with surrounding DM. This process could lead to observable SM signals with the aid of specific force mediators. The signal's energy spectrum depends on the DS processes and mediators involved, correlating strongly with BH properties like mass and number density.

Since the BH temperature and evaporation rate scale with the inverse of its mass, hotter, lighter BHs will have evaporated by today and thus cannot be present in large numbers.\footnote{A small population of lighter and short-lived {\it young} BHs may be present today as the remnant from the continuous evaporation of an older and heavier population. We perform a comprehensive study of the abundance of light BHs and their observational signatures in an upcoming paper~\cite{futurework}.}
This translates into a lower bound on the BH mass for it to have survived until today; for non-rotating BHs it is equal to $M_\BH \gtrsim 5 \times 10^{14}~\gr$ and corresponds to a BH temperature of $T_\BH \approx 21~\MeV \, \bl( 5 \times 10^{14}~\gr / M_\BH \br)$ and a BH lifetime of $\tau_\BH \approx 16~\Gyr \, \bl( M_\BH / 5 \times 10^{14}~\gr \br)^3$, roughly the age of the Universe \cite{Carr:2020gox}. The particle emission rate due to the Hawking radiation of such BHs is roughly $\Gamma_\BH \sim 6 \times 10^{21}~\sec^{-1} \, \bl( 5 \times 10^{14}~\gr / M_\BH \br)$. Black holes with masses around these scales cannot have an astrophysical origin and therefore must be formed in the very early Universe, hence their name {\it Primordial Black Holes} (PBHs).
Formation of PBHs can involve the gravitational collapse of overdensities originating from inflation, cosmic strings, domain walls, bubble collisions, dissipative dark sectors, and other exotica (\cite{Carr:2020gox,Gurian:2022nbx} and references therein). Note that PBHs can themselves be a DM constituent, although they are extended objects and not particles. The DM mass fraction $f_\PBH$ that is found in PBHs can be constrained from the absence of observable Hawking radiation and of CMB spectral distortions (constraining $M_\PBH$ in the asteroid mass range, between $5 \times 10^{14}~\gr$ and $10^{17}~\gr$), and from lensing (for masses above $10^{23}~\gr$); for a review of constraints see Ref.~\cite{Carr:2020gox} and references therein. While the window between $10^{17}~\gr$ and $10^{23}~\gr$ is currently unconstrained and $f_\PBH = 1$ is allowed, future probes such as the gamma-ray surveys AMEGO-X \cite{Caputo:2022xpx} and e-ASTROGAM \cite{e-ASTROGAM:2016bph}, and dedicated microlensing searches for M31 (the Andromeda galaxy), can narrow it further at its lower and higher mass ends, respectively \cite{Drlica-Wagner:2022lbd}.

\begin{figure}
\centering
\includegraphics[width=0.5\linewidth]{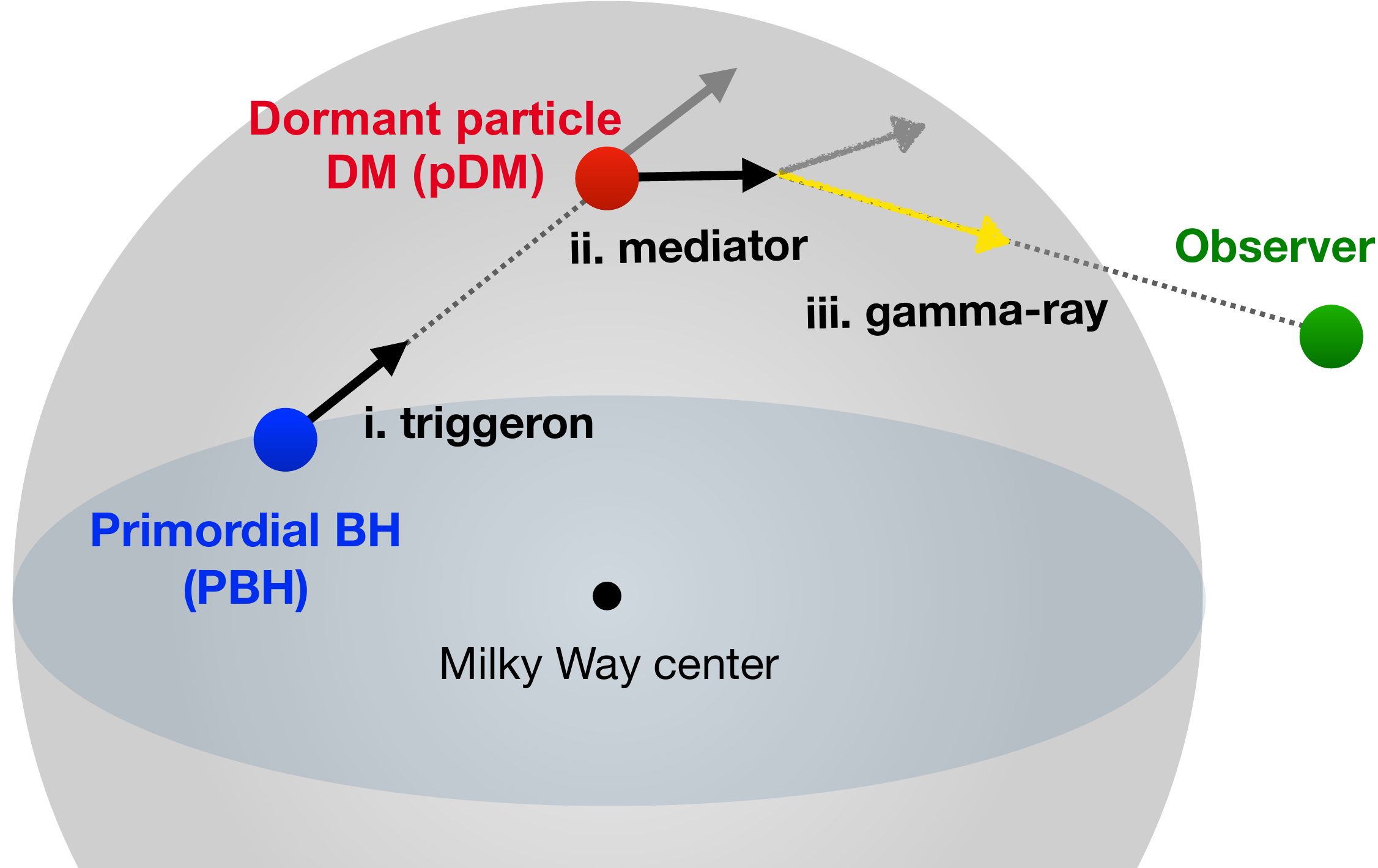}
\caption{The geometry of PBH-induced dark sector indirect detection signals, illustrating the chain of processes listed in \Eq{eq:steps_general}. The gray sphere represents the region-of-interest of the DM halo. The PBHs located at the blue dot emit triggerons via Hawking radiation ({\it i}). These travel for some distance before annihilating with ambient pDM at the red dot ({\it ii}). The DM--SM mediators produced in these annihilations promptly decay into photons, producing a gamma-ray signal ({\it iii}).}
\label{fig:geometry2}
\end{figure}

Thus, the possibility of dark complexity presents us with the following intriguing scenario: if (part of) the DM is made of particles belonging to a broader DS, with its own interactions and portals to the SM, a large enough number of sufficiently hot PBHs will act as a ``catalyst'' relative to the otherwise dormant particle DM (pDM) ``reactant'';\footnote{Once again, note that the catalyst PBHs and the reactant DM {\it particles} both contribute to the {\it total} dark matter density and are thus part of what we mean by ``dark matter''. We denote the particle DM component by pDM, in order to distinguish it from the {\it total} DM (\ie, particles + black holes).} the flux of triggeron DS particles emitted by the PBHs Hawking radiation will interact with the pDM and trigger DS processes that may lead to an indirect detection signal. 
As illustrated in Fig.~\ref{fig:geometry2}, the steps that lead to the indirect detection signal can be schematically represented by the following chain of processes 
\bea\label{eq:steps_general}
    \text{(i)  emission: } &&\!\!\!\!\!\!\! \text{ PBH } \to \text{ triggeron}\,\, \text{ (via Hawking radiation);} \nonumber\\
    \text{(ii) annihilations: } && \!\!\!\!\!\!\! \text{ triggeron } + \text{ pDM } \to \text{ mediator } + \text{other} \ ; \nonumber\\
    \text{(iii) decays: } && \!\!\!\!\!\!\! \text{ mediator } \to \text{ SM } + ... \ .
\eea

In the present work, we explore the observability of these (p)DM indirect detection signals triggered by the DS Hawking radiation of PBHs. For concreteness, we focus on galactic photon signals in the gamma-ray spectrum, and on asteroid-mass PBHs. As indicated earlier, this mass and energy range is singled out both by the longevity of the PBHs in question ($M_\PBH \gtrsim 5 \times 10^{14}~\gr$ in order for the PBHs to live longer than the age of the Universe), the sensitivity of upcoming gamma-ray surveys such as AMEGO-X (energies $E_\gamma \gtrsim 0.1~\MeV$ \cite{Caputo:2022xpx}), and the typical photon energy from Hawking radiation for this PBH mass range ($E_\gamma \approx 5.8~T_\PBH$, PBH masses of $M_\PBH \lesssim 6 \times 10^{17}~\gr$). As a proof of concept, we consider two models that give rise to such signals: One in which the pDM is composed of asymmetric particles, and the triggerons coming from the PBHs are its antiparticles. The second in which the pDM is the lowest-energy state of a multiplet, and the triggerons are dark gauge bosons that convert the pDM into a higher DS state. These higher DS states subsequently decay back into the DM particle and a photon (\eg, via an inelastic dipole transition).

This paper is organized as follows: In Section \ref{sec:processes}, we discuss in detail how DS processes are triggered in the presence of PBHs and compute the differential photon flux to which they give rise. Section \ref{sec:models} is devoted to concrete examples of the models mentioned in the previous paragraph, and to exploring the observability of their associated photon signal. We conclude in Section \ref{sec:conclusions}. In addition, there are three appendices dealing with topics of interest: the possibility of DM halo evaporation and of triggeron annihilation in the vicinity of the PBHs (Appendix \ref{app:A}), the detailed derivation of the photon signal in these models (Appendix \ref{app:B}), and the details of the dipole transition model (Appendix \ref{app:C}).

\section{Triggering: Processes and Signal}
\label{sec:processes}

In this section, we discuss in detail the central idea of our paper, namely, the mechanism through which the Hawking radiation triggerons emitted by PBHs interact with ambient DM particles to produce an indirect detection signal. Of necessity, we consider a ``mixed DM'' scenario in which the DM is comprised of at least two different populations, the PBHs and some kind of particle DM or pDM that has significant interactions with the triggerons. While other DM species may be present, we focus only on PBH and pDM populations inside the galaxy. As shown in Ref.~\cite{Ray:2021mxu}, the Hawking radiation of photons from an extragalactic population of PBHs is subdominant to the one coming from the galactic population for most of the photon energy spectrum. Therefore, the Hawking triggeron flux from galactic PBHs is also larger than the one produced by extragalactic ones. For simplicity, we only consider a monochromatic mass distribution of non-rotating PBHs. As a result, the $i$-th DM population (pDM or PBHs) can be parametrized by the mass $m_i$ of its constituents (namely, the particle mass $m_\pDM$, or the PBH mass $M_\PBH$) and the mass fraction $f_i$ of the total halo mass that is in the $i$-th species. As we shall see, $f_\PBH$ need not be particularly large in order to generate an interesting and observable DS signal, although unsurprisingly $f_\pDM$ needs to be significantly larger.

Finally, and again for simplicity, we assume that the triggeron--pDM annihilations occur via 2-to-2 scatterings, and consider only two-body decays of the mediators. More concretely, the chain of events in \Eq{eq:steps_general} becomes
\bea\label{eq:steps}
    \text{(i) emission: } &&\!\!\!\!\!\!\! \text{ PBH } \to \text{ triggeron ;} \nonumber\\
    \text{(ii) annihilations: } && \!\!\!\!\!\!\! \text{ triggeron } + \text{ pDM } \to \text{ mediator } + X \ ; \nonumber\\
    \text{(iii) decays: } && \!\!\!\!\!\!\! \text{ mediator } \to \gamma + Y \ ,
\eea
where $X$ and $Y$ are additional particles in these processes, with masses $m_X$ and $m_Y$, respectively. The identities of $X$ and $Y$ depend on the details of the DS. For example, $X$ could be an additional mediator and $Y$ an additional photon, or they could be something else entirely. In this section, we discuss the steps in \Eq{eq:steps} in a model-independent fashion; we defer the study of specific models and their signatures to \Sec{sec:models}. There are some potential issues with this scenario, such as the evaporation of the DM halo due to triggeron--pDM annihilations, and the impact that a pDM population gravitationally bound to PBHs (called DM ``spikes'' or ``mini-halos'' in the literature) can have on the signal. We postpone their discussion until \App{app:A}, where we find these issues to be ultimately negligible.

\subsection{Kinematics}

Let us begin by considering the kinematic requirements that must be satisfied in order for the triggering of pDM. Triggerons are emitted by PBHs as Hawking radiation. A non-rotating black hole (primordial or not) of mass $M_\BH$ has a temperature \cite{Hawking:1974rv,Hawking:1975vcx}
\beq\label{eq:hawk_temp}
    T_\BH = \frac{1}{8\pi G_N M_\BH} \ ,
\eeq
where $G_N$ is the gravitational constant. The corresponding emission rate spectrum per degree of freedom of a Hawking radiation particle of mass $m$ and spin $s$ is given by
\beq\label{eq:hawk_rad}
    \frac{{\rm d}^2 N}{{\rm d}t \,{\rm d}E} = \frac{1}{2 \pi} \frac{\Gamma_s(E, m; M_\BH)}{e^{E/T_\BH} - (-1)^{2s}} \ ,
\eeq
where $\Gamma_s(E, m; M_\BH)$ is known as the {\it graybody factor}, and it depends on the properties of the radiated particles. In the high-energy limit ($G_N M_\BH E \gg 1$), however, it reduces to $\Gamma_s = 27 G_N^2 M_\BH^2 E^2$; the emission power $P = \int \dd E \, E \, \frac{{\rm d}^2 N}{{\rm d}t {\rm d}E}$ obtained with such a simplified factor is that of a blackbody.

Equation (\ref{eq:hawk_rad}) shows that, crucially, the triggeron must be sufficiently light to be emitted as Hawking radiation by the PBHs. Indeed, if the triggeron mass $m_\trg$ is much larger than temperature $T_\PBH$, then the triggeron flux is Boltzmann-suppressed and therefore negligible. Triggerons with masses comparable to $T_\PBH$ receive corrections to $\Gamma_s$ for producing non-relativistic particles, altering the low-energy spectrum of the triggering production~\cite{Cui:2024uwk}. 

Another requirement involves the triggeron energy $E_\trg$, which must be large enough for the second step in \Eq{eq:steps} (\ie, the scattering process $\text{triggeron} + \text{pDM} \to \text{mediator} + X$) to be kinematically allowed. Therefore, the actively radiating PBHs in the asteroid-mass range are naturally suited for probing dynamics of MeV--GeV dark sectors.

In summary, the triggering process presupposes that
\bea
    m_\trg & \lesssim & E_{\rm pk} \sim \mathcal{O}(\text{few}) \times T_\PBH \ , \label{eq:kin1} \\
    E_\trg & \geq & \Delta m + m_X \ , \label{eq:kin2}
\eea
where $E_{\rm pk}$ is the peak energy of the Hawking radiation spectrum and $\Delta m \equiv m_\med - m_{\pDM}$ is the mass splitting between the DS--SM mediator and the DM particle, ignoring the subdominant contribution from the initial pDM kinetic energy.

\subsection{Indirect Detection Signal}

As stated in the introduction, the dormant pDM constituent is awakened by the triggeron emitted by the PBHs through Hawking radiation. These triggeron--pDM interactions produce the mediators which subsequently decay into SM particles; photons in our case. Na\"ively, the resulting photon flux $\Phi_\gamma$ is just like any indirect detection signal, depending on the particle properties of the mediator, such as its decay rate $\Gamma_\med$, its mass $m_\med$, and its abundance $\rho_\med = f_\med \rho_\DM$ in the halo: $\Phi_\gamma \propto \Gamma_\med f_\med / m_\med$. However, since the mediator population is not part of the original DM halo but instead arises dynamically from the PBH-induced triggeron--pDM interactions, the amplitude of this indirect detection signal ultimately depends on the PBH population. Indeed, if each triggeron--pDM collision results in a number $\overline{c}_\med$ of mediators, their total number $N_\med$ in the halo depends on its decay rate and its production rate:
\bea
    \dot N_\med & = & - \Gamma_\med N_\med + \frac{\overline{c}_\med}{M_\PBH} \mathcal{P}_\ann \Gamma_\trg f_\PBH M_{\rm halo} \ , \\
    \Rightarrow f_\med & \simeq & \frac{\overline{c}_\med m_\med}{M_\PBH} \frac{\Gamma_\trg}{\Gamma_\med} f_\PBH \bl( 1 - e^{-\Gamma_\med t} \br) \ ,
\eea
where $M_{\rm halo}$ is the total DM mass in the halo, $\mathcal{P}_\ann$ the total probability that a triggeron has of annihilating with ambient pDM, and $\Gamma_\trg$ the total triggeron emission rate by one PBH which is obtained by applying \Eq{eq:hawk_rad} to the triggeron and integrating over its energy, $\Gamma_\trg = \int\dd E_\trg \ {\rm d}^2 N_\trg^{1\PBH} / {\rm d}t {\rm d}E_\trg$. In the last equality above, we have taken $\mathcal{P}_\ann \approx 1$ and solved for $f_\med \equiv m_\med N_\med / M_{\rm halo}$ assuming a negligible initial mediator population. In other words, if the mediator's lifetime is shorter than the age of the galaxy (\ie, $\Gamma_\med t \gg 1$) its population in the halo is in a steady state, being constantly depleted by its decays and replenished by the PBH production. As a result, the indirect detection signal scales like $\Phi_\gamma \propto \Gamma_\med f_\med / m_\med = \overline{c}_\med  \Gamma_\trg f_\PBH / M_\PBH$. The scaling is an interesting consequence of DM triggering: the details of the DS interactions do not impact the amplitude of the indirect detection signal they produce, in the limit when $\mathcal{P}_\ann \approx 1$. While these details determine the photon energy distribution (\ie, the {\it shape} of the spectrum), the amplitude of the spectrum depends solely on the PBH properties. This suggests that, were the Hawking radiation of PBHs to be detected, it could in principle be accompanied by a DS indirect detection signal of similar size.

Let us now devote our attention to the resulting photon signal, the study of which is the subject of \App{app:B}, in the most general case.
Given the discussion in the introduction and the previous sections, we focus on scenarios where triggerons have a high probability of annihilating with ambient pDM, specifically when $\mathcal{P}_\ann \approx 1$. In contrast, feeble interactions between pDM and triggerons yield no signal, effectively leaving pDM inert. Following in the footsteps of Ref.~\cite{Agashe:2020luo}, we can parametrize the probability of triggeron--pDM annihilations in terms of the {\it annihilation coefficient} $\Lambda$, defined as the inverse triggeron mean-free-path in units of the typical halo length scale:
\bea
    \Lambda \equiv \lambda_\mfp^{-1} r_s & = & n_{\pDM,0} \, \sigma_\ann r_s = \frac{\rho_0f_\pDM  \sigma_\ann r_s}{m_\pDM}
     =  \bl( \frac{f_\pDM \sigma_\ann / m_\pDM}{\bl( 22~\MeV \br)^{-3}} \br) \bl( \frac{\rho_0}{0.838~\GeV \cm^{-3}} \br) \bl( \frac{r_s}{11~\kpc} \br) \ ,
    \label{eq:Lambda}
\eea
where $\sigma_\ann$ is the triggeron--pDM annihilation cross section, $n_{\pDM,0} = (\rho_0f_\pDM )/ m_\pDM$, and $\rho_0$ and $r_s$ are the DM density and scale radius parameters describing the total DM density halo profile $\rho_\DM(r)$. We have used as benchmarks the values for the Navarro-Frenk-White (NFW) profile \cite{Navarro:1994hi,Navarro:1995iw,Navarro:1996gj}, which has $\rho_\DM(r) = \rho_0 \eta(r/r_s)$, with $\eta(x) \equiv 1/x(1+x)^2$, $\rho_0 = 0.838~\GeV \cm^{-3}$, and $r_s = 11~\kpc$ \cite{deSalas:2019pee}. Here and everywhere in this paper we assume that the $i$-th DM species, of mass fraction $f_i$, follows the same universal profile, and therefore $\rho_i(r) = f_i \rho_\DM(r)$. Note that in general $\Lambda$ depends on the triggeron energy $E_\trg$ through $\sigma_\ann$. We are interested in the case where the triggeron mean-free path is very short compared to the typical galactic scales, \ie, $\Lambda \gg 1$. This means that $f_\pDM \sigma_\ann / m_\pDM \gg \bl( 22~\MeV \br)^{-3} = 20~\cm^2/\gr$, which implies a value for $\sigma_\ann / m_\pDM$ larger than the Bullet Cluster constraint on the DM self-interaction cross section, $\sigma_{\rm self} / m_\pDM \approx 1~\cm^2 / \gr$ \cite{Markevitch:2003at,Randall:2008ppe}. While these are cross sections for two different processes (one is triggeron--pDM annihilations, and the other pDM--pDM scatterings), one must keep this bound in mind when considering specific DS models. A way to avoid this constraint is to require the pDM to make up only a fraction $f_\pDM < \mathcal{O}(\text{few }10\%)$ of the total DM \cite{Robertson:2016xjh}. In the case where the annihilation coefficient $\Lambda$ is very large, the triggeron--pDM annihilations take place locally, \ie, roughly in the location of the PBHs radiating the triggerons. If the mediator lifetime is also very short, which we assume to be the case, the photon signal coming from these dark processes is also local.

In \App{app:B3} we show that in this $\Lambda \to \infty$ {\it local limit}, the differential photon flux along a line-of-sight distance (los) $\ell$ to a region-of-interest (ROI) of solid angle $\Omega_{\rm ROI}$ is given by
\bea
    \frac{{\rm d} \Phi_\gamma}{{\rm d} E_\gamma} & = & \frac{1}{4 \pi} \frac{f_\PBH}{M_\PBH}  \, J_\dec \, \int \dd E_\trg \ \frac{{\rm d}^2 N^{\rm 1 PBH}_\trg (E_\trg)}{{\rm d}t \,{\rm d}E_\trg} \, \frac{{\rm d} N_\gamma(E_\trg, E_\gamma)}{{\rm d} E_\gamma} \ , \label{eq:dPhidE_simplest} \\
    \text{where} \quad J_\dec & \equiv & \int\limits_{\rm ROI} \dd\Omega \int\limits_{\rm los} \dd \ell \ \rho(r) \label{eq:Jdec}
\eea
is the $J$-factor for decaying DM common in the indirect detection literature. In the photon flux calculation, the $J$-factor is further re-scaled by $f_\PBH$ to describe the ``decaying'' (\ie, evaporating) PBH that source the signal. For reference, the $J$-factor for a ROI spanning an angle $\theta_{\rm ROI}$ around the galactic center (GC) is given by \Eq{eq:Jfac_cone}. In our subsequent numerical analyses we consider an ROI with $\theta_{\rm ROI} = 5\degree$, an NFW profile with $\rho_0$ and $r_s$ as listed in the previous paragraph, and a distance $d_\odot = 8.122~\kpc$ from the Sun to the GC \cite{GRAVITY:2018ofz}, which together give $J_\dec \approx 3.8 \times 10^{24}~\MeV \cm^{-2}$.

The quantity ${\rm d}^2 N_\trg^{1\PBH}(E_\trg)/{\rm d}t {\rm d}E_\trg$ in \Eq{eq:dPhidE_simplest} is the emission rate spectrum of the triggerons emitted by one PBH, and is given by \Eq{eq:hawk_rad}. We compute this spectrum using graybody factors obtained from the publicly available \href{https://blackhawk.hepforge.org/}{\tt BlackHawk} program \cite{Arbey:2019mbc, Arbey:2021mbl}. ${\rm d} N_\gamma (E_\trg, E_\gamma) / {\rm d}E_\gamma$ is the photon spectrum resulting from the triggeron--pDM annihilations and subsequent mediator decays, and it is of the following general form:
\beq\label{eq:dNgamma_general}
    \frac{{\rm d} N_\gamma (E_\trg, E_\gamma)}{{\rm d} E_\gamma} = \int\dd E_\med \ \frac{{\rm d} N_\med (E_\med, E_\trg)}{{\rm d} E_\med} \frac{{\rm d} N_\gamma^{1\dec}(E_\gamma, E_\med)}{{\rm d} E_\gamma} \ ,
\eeq
where ${\rm d} N_\med (E_\med, E_\trg)/{\rm d} E_\med$ is the mediator energy spectrum per annihilation event, and ${\rm d} N_\gamma^{1\dec}(E_\gamma, E_\med) / {\rm d} E_\gamma$ is the photon spectrum per decay event, for a given mediator energy. The ${\rm d} N_\gamma / {\rm d}E_\gamma$ photon spectrum peaks at an energy $E_{\gamma*} = (m_\med^2 - m_Y^2)/2 m_\med$ corresponding to the the photon energy in the mediator's rest frame, but it is broadened because in the galactic frame the mediators come out of the annihilation point with a boost factor that depends on $E_\trg$ and their direction of motion. The result is a photon spectrum loosely trapezoidal in shape, whose precise form depends on the energy of the particles in the center-of-mass frame of the triggeron--pDM annihilations, the energy of the particles in the mediator's rest frame, and the boosts of these frames with respect to the galactic frame. The precise form of the photon spectrum is model-dependent; the top panels of Figs.~\ref{fig:badm_spectra} and \ref{fig:dtdm_spectra} show two examples of this spectrum for the two models we discuss in detail in \Sec{sec:models}.

The reason behind this shape can be intuitively understood as follows. While in the mediator's rest frame, the photon is monochromatic (for the two-body decays we consider), in the galactic frame its energy is distributed according to a box-shape function {\it for a fixed mediator energy}. The width of this box is determined by the mediator's energy. This box-shaped distribution is in fact due to the mediator's rest frame moving with respect to the galactic frame, and therefore the photons can be emitted in all directions with respect to the direction of this motion, including parallel or antiparallel. The same is true, {\it mutatis mutandis}, of the mediator energy distribution: In the galactic frame, the mediators themselves can come out at any angle with respect to the direction of motion of the triggeron--pDM center-of-mass frame relative to the galactic center. Thus, the width of the box-shaped photon distribution depends on the mediator energy, which itself is drawn from a box-shaped mediator distribution whose width depends on the triggeron energy. The result is an integral of box distributions (the photon's for a fixed mediator energy, and the mediator's for a fixed triggeron energy), which amounts qualitatively to a series of ``stacked`` photon boxes of varying width. The overall result is a trapezoidal distribution for the photons. The exact formula for this spectrum ${\rm d}N_\gamma / {\rm d}E_\gamma$ is given in \Eq{eq:photon_spectrum}; for its derivation see \App{app:B2}.

For finite $\Lambda$, the local limit ceases to be exact and many complications arise which prevent a straightforward calculation of the differential photon flux. The main difficulty is that, since the triggerons can now travel finite galactic distances, their annihilations do not necessarily take place in the same location as their emission from the PBHs; see \App{app:B1}. This results in a non-isotropic photon emission at each point in the halo, with a complicated interdependence between the triggeron energy, the photon angular distribution, and the geometry of the ROI. In other words, the differential photon flux cannot be factorized into various independent pieces separately encoding the astrophysics and particle physics, as done in \Eq{eq:dPhidE_simplest}. The non-local effects consequence of a finite $\Lambda$ were first studied in Ref.~\cite{Agashe:2020luo}, albeit in the context of multi-stage DM annihilations instead of Hawking evaporation of PBHs. However, in that paper's model, the outgoing photons were highly collimated due to the mediators having a large boost, which greatly simplifies the calculations. This is not necessarily true in our case, because the boost of the mediators depends on the triggeron energy, which in turn follows a Hawking radiation distribution. We direct the reader to \App{app:B} for a derivation of the differential photon flux. Nevertheless, one can estimate the impact that a finite $\Lambda$ has on the signal by multiplying it by a suppression factor $\mathcal{J}(\Lambda)$ that depends on the geometry of the ROI and the annihilation coefficient $\Lambda$. The updated photon differential flux can therefore be approximated as
\beq
    \frac{{\rm d} \Phi_\gamma}{{\rm d} E_\gamma} & \simeq & \frac{1}{4 \pi} \frac{f_\PBH}{M_\PBH}  \, J_\dec \, \int \dd E_\trg \ \frac{{\rm d}^2 N^{\rm 1 PBH}_\trg (E_\trg)}{{\rm d}t \,{\rm d}E_\trg} \, \frac{{\rm d} N_\gamma^{1\dec}(E_\trg, E_\gamma)}{{\rm d} E_\gamma} \, \mathcal{J}(\Lambda(E_\trg)) \ . \label{eq:dPhidE_curlyJ}
\eeq
The exact functional shape of $\mathcal{J}(\Lambda)$ depends on the details of the ROI and the distances the triggerons are able to travel before annihilating. Independently of any particulars, however, we have $\mathcal{J}(\Lambda) \to 1$ as $\Lambda \to \infty$ in the local limit, and $\mathcal{J}(\Lambda) \to 0$ as $\Lambda \to 0$. In \App{app:B4}, we show how we can find the typical length scale associated with a given ROI, and how that can be used to estimate $\mathcal{J}(\Lambda)$. For example, we find that $\mathcal{J}(1) = 0.67$, while $\mathcal{J}(4) = 0.96$, for a ROI $5\degree$ around the GC.

Note the dependence of $\mathcal{J}(\Lambda)$ on the triggeron energy $E_\trg$ through $\sigma_\ann$, which means that it cannot in general be pulled out of the integral. The most representative value of $\Lambda$ is when $E_\trg = E_{\rm pk}$ (peak energy of the Hawking radiation spectrum), which according to ${\rm d}^2 N_\trg^{1\PBH} / {\rm d}t {\rm d}E_\trg$ is the most common triggeron energy; annihilation events with $E_\trg \neq E_{\rm pk}$ contribute less to the overall photon signal. As is typical of cross sections, larger $E_\trg$ translates into smaller $\sigma_\ann$ and thus smaller $\Lambda$. As discussed in more detail in \App{app:B4}, however, this energy dependence has an impact only on the higher- and lower-energy ends of the photon differential flux, since larger $E_\trg$ means larger boosts for the mediators, and thus more extreme photon energies. In other words, the bulk of the photon spectrum comes from triggerons with $E_\trg \approx E_{\rm pk}$, and one could therefore further simplify \Eq{eq:dPhidE_curlyJ} by assuming a constant $\mathcal{J}(\Lambda)$ defined at $E_\trg = E_{\rm pk}$.

Throughout this paper we work mostly in the local limit, \ie, with \Eq{eq:dPhidE_simplest}. As already stated, this is both the simplest case, and the most phenomenologically interesting, since the premise of our present study is precisely the existence of a DS where the DM has sizable interactions with other dark particles. As we show in \Sec{sec:models}, we can construct several DS toy models where $\Lambda \gg 1$ and the local limit is a good approximation, with $\mathcal{J}(\Lambda)$ yielding a suppression in the flux of only $\sim \mathcal{O}(1\%)$ or $\mathcal{O}(10\%)$ relative to the exactly local limit.

\subsection{Indirect Detection with AMEGO-X}\label{subsec:amegox}

Future gamma-ray surveys such as AMEGO-X \cite{Caputo:2022xpx} and e-ASTROGAM \cite{e-ASTROGAM:2016bph} can potentially observe the DS signals discussed in this work. Here, we focus on the AMEGO-X telescope as a convenient benchmark to test the sensitivity of future indirect detection experiments to the DS spectral feature in the Hawking radiation spectra. AMEGO-X observations will measure gamma-rays from about $0.1~{\rm MeV}$ to $1~{\rm GeV}$, which overlaps with the Hawking temperature of asteroid-mass PBHs.

We use a likelihood analysis method in order to obtain the sensitivity of the future AMEGO-X MeV indirect detection experiment for our benchmark models. We first calculate the expected photon number in the $i$-th energy bin with a binning choice of $50$ equal-width bins on logarithmic scale from $0.2~{\rm MeV}$ to $300~{\rm MeV}$,
\bea
\sigma_i=\int_{E_i^{\rm min}}^{E_i^{\rm max}} {\rm d}E_\gamma \, A(E_\gamma) \, T_{\rm obs} \, \bl( \frac{{\rm d}\Phi_\gamma}{{\rm d}E_\gamma} + \frac{{\rm d}\Phi_{\gamma, \rm bkg}}{{\rm d}E_\gamma} \br) \ .
\eea
Here ${E_i^{\rm min}}$ and ${E_i^{\rm max}}$ are the edges of the $i$-th bin. The effective area $A(E_\gamma)$ as a function of gamma-ray energy is obtained from Ref.~\cite{Caputo:2022xpx} for untracked Compton events, tracked Compton events, and pair events. We do not include single-site events because their photon directions cannot be reconstructed to reduce the background for the GC search. We choose a region of interest of $5\degree$ about the Galactic Center and an observation time of $T_{\rm obs}=10^{8}~{\rm s}$. The DS signal is combined with the background contribution $\Phi_{\gamma, \rm bkg}$ for the three types of AMEGO-X events under consideration, as reported in~\cite{Caputo:2022xpx}, to produce the total gamma-ray flux.
In order to determine whether two models are distinguishable, we assume a particular ``true'' model from which an observed gamma-ray signal originates and produces a predicted binned signal count, $n_i$. We compare this count with those predicted by a ``test'' model to determine the likelihood of the counts originating from the ``test'' model. The likelihood that the signal from the ``true'' model could originate from the ``test'' model is
\bea
L = \exp{\left(\sum_i n_i \ln{\sigma_i}-\sigma_i-\ln{n_i!}\right)} \ .
\label{eq:likelihood}
\eea
The corresponding level of significance $\Sigma$ of comparing the likelihood of the ``true'' model $L_{\rm true}$ and a ``test'' model $L$ is therefore given by
\bea
    \Sigma^2 = - 2 \ln{\left(\frac{L}{L_{\rm true}}\right)} \ .
\label{eq:TS}
\eea
In the following sections, we use the likelihood significance $\Sigma$ to examine the discovery power of future gamma-ray observations to identify the capability of observing the triggered indirect detection signals (distinguishable from a background composed of a monochromatic SM-emitting PBH scenario) as well as being able to distinguish between different models. We define a threshold of $\Sigma>2$ when declaring two signals as distinct. For computational purposes, when searching through a model's parameter space, we follow the approach detailed in \cite{Agashe:2022jgk,Agashe:2022phd} which assumes that the likelihood contains no local minima and follow a binomial search pattern.

\section{Triggering: Dark Sector Models}
\label{sec:models}

Let us now turn our attention to examples of dark sector models that give rise to the triggered indirect detection photon signal described in the previous section. As stated in the introduction, while the existence of a signal like this could be a rather generic feature of dark complexity, its characteristics depend on the precise nature of the dark sector interactions, of the dark particles involved, and of their portal to the SM. Nevertheless, as we will demonstrate shortly, such a signal may in principle be observed by upcoming experiments.

In this section we consider two templates of the triggered dark matter photon signal and, as a proof of principle, construct two dark sector toy models that give rise to these signals. In the first model, the triggerons are anti-DM particles which, upon annihilating with the ambient particle DM, produce an axion-like particle (ALP) mediator which subsequently decays into two photons, giving rise to a broad spectrum. In the second model, the triggerons scatter inelastically with the pDM, converting it into a heavier partner. This heavier dark state, the mediator, decays back into the original pDM and a single photon; the associated signal is a line-like spectrum. Based on these models we then show how their signals can be detected by the proposed AMEGO-X experiment.

\subsection{Broad spectrum: Boosted anti-DM (BaDM) model}\label{subsec:badm}

\begin{figure}
\centering
\includegraphics[width=0.7\linewidth]{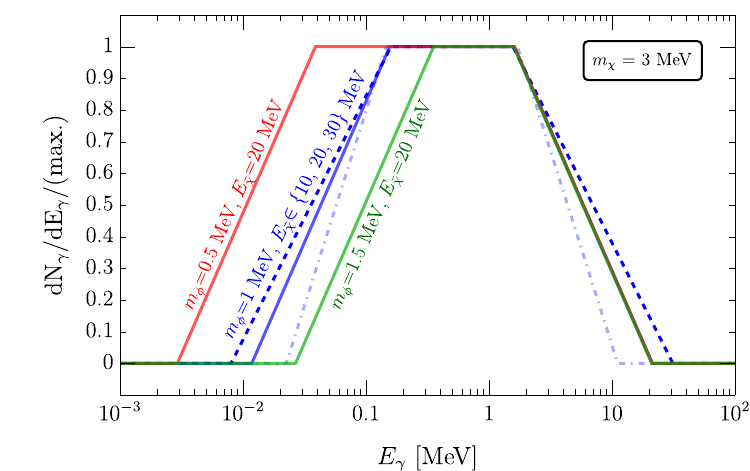}
\\
\includegraphics[width=0.7\linewidth]{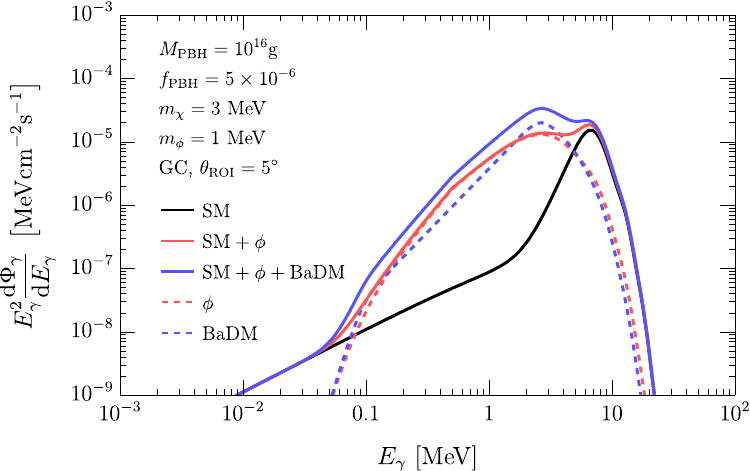}
\caption{{\it Top:} The trapezoid-shaped spectrum $dN_\gamma/dE_\gamma$ of \Eq{eq:dNgamma_general}, for the BaDM model, and various choices of the triggeron energy $E_\trg = E_{\anti{\chi}}$ and the mediator mass $m_\med = m_\phi$. We have taken $m_\chi = 3~\MeV$. For convenience, we have normalized the spectra such that their maximum value is always $1$. The blue dot-dashed, solid, and dashed spectra correspond to $m_\phi = 1~\MeV$ but $E_{\anti{\chi}} = 10~\MeV$, $20~\MeV$, and $30~\MeV$, respectively. The red and green lines both have $E_{\anti{\chi}} = 20~\MeV$, but $m_\phi = 0.5~\MeV$ and $1.5~\MeV$, respectively. Note how a larger triggeron energy $E_{\anti{\chi}}$ results in more boosted photons, and thus a broader spectrum. Increasing $m_\phi$ changes the center of the spectrum (which is $E_{\gamma*} = m_\phi/2$), and also narrows down the distribution, since a larger mediator mass $m_\phi$ means that its RF is less boosted for a fixed triggeron energy $E_{\anti{\chi}}$. {\it Bottom:} The gamma-ray signal (photon spectral flux density times energy) in the BaDM model, coming from a ROI of $5\degree$ around the GC of the Milky Way. The black line is the expectation from the SM (\ie, direct Hawking radiation of photons plus final-state radiation contributions). The red dashed line is the contribution to the signal from direct Hawking emission (and subsequent decays) of the $\phi$ mediators by the PBHs. The blue dashed line is the contribution from ambient $\chi$ pDM triggered by Hawking radiation by PBHs of $\anti{\chi}$ triggerons, following \Eq{eq:steps_badm}. The solid lines are the sum of SM and these contributions. The blue solid line is what the observer ultimately sees. For all curves, we have chosen the benchmark parameter values of $M_\PBH = 10^{16}~\gr$, $f_\PBH = 5 \times 10^{-6}$ ($1/100$ of the current bound for this mass \cite{Carr:2020gox}), $m_\chi = 3~\MeV$, and $m_\phi = 1~\MeV$. We computed the photon differential flux in the local limit ($\Lambda \gg 1$).}
\label{fig:badm_spectra}
\end{figure}

For our first toy model, we consider an asymmetric DM scenario based on that of Ref.~\cite{Agashe:2020luo} which gives rise to a broad photon spectrum. The dark sector consists of two fields, a scalar DM $\chi$ charged under a dark global $U(1)_d$ symmetry, and an ALP mediator $\phi$. The lagrangian density of this DS contains the interactions
\beq\label{eq:badm_lagrangian}
    \mathcal{L}_{\rm DS} \supset y \abs{\chi}^2 \phi^2 + \lambda \abs{\chi}^4 - \frac{1}{4} g_{\phi \gamma \gamma} \phi F_{\mu\nu} \tilde{F}^{\mu\nu} \ ;
\eeq
where $F_{\mu\nu}$ is the SM electromagnetic tensor. The $\anti{\chi}$ antiparticles are absent from the ambient DM due to an unspecified process in the early Universe in which an asymmetry in the dark $U(1)_d$ charge was generated. However, the $\anti{\chi}$s can be emitted by PBHs as triggerons, with a boost factor of $E_\trg / m_\trg = E_{\anti{\chi}} / m_\chi \sim T_\PBH / m_\chi$. The $\anti{\chi}$ then annihilate with ambient $\chi$ pDM to produce a pair of $\phi$ mediators. For sufficiently large $E_{\anti{\chi}}$, the $\anti{\chi}\chi$ annihilations can be energetic enough to produce two $\phi$ mediators even if they are heavier than the $\chi$s; nevertheless for simplicity, we always assume $m_\phi < m_\chi$. The discussion around \Eq{eq:dNgamma_general} can now be applied to this scenario. While the photons are isotropically distributed and monochromatic in the $\phi$ rest frame (RF) with energy $E_{\gamma*}^{[\mathrm{RF}]} = m_\phi/2$, in the galactic frame (GF), their spectrum is broadened into a trapezoidal distribution $d N_\gamma / dE_\gamma$ due to the distribution of the boosts of the $\phi$s' RFs relative to the GF.\footnote{In this model, and throughout this paper, we assume the annihilation products are isotropically distributed in the CF of the annihilation event. This is true if there is no net spin in the collisions. Assuming the spins of the triggerons and the pDM are randomly distributed in every point of the galaxy, this is indeed the case. See \App{app:B} for more details.} The center of the spectrum remains at $E_{\gamma*} = m_\phi/2$. In the top panel of \Fig{fig:badm_spectra} we show the photon spectrum $d N_\gamma / dE_\gamma$, for given $E_\trg$, $m_\chi$, and $m_\phi$.

In summary, for this {\it ``boosted anti-dark matter''} (BaDM) model, the chain of processes in \Eq{eq:steps} is
\bea\label{eq:steps_badm}
    \text{BaDM model:} \quad 
    \begin{cases}
        \text{emission: } \quad & \text{PBH } \to \anti{\chi} \ ;\\
        \text{annihilations: } & \anti{\chi} \chi \to \phi \phi \ ;\\
        \text{decays: }  \quad & \phi \to \gamma \gamma \ ,
    \end{cases}
\eea
with the following kinematic requirements
\beq
    m_\phi < m_\chi \lesssim \mathcal{O}(\text{few}) \times T_\PBH \ .
\eeq

The bottom panel of \Fig{fig:badm_spectra} shows the photon signal coming from the triggered DM of our BaDM model according to \Eq{eq:dPhidE_simplest} for values of the model parameters that yield a signal observable by AMEGO-X and in the local limit of the signal ($\Lambda \gg 1$) where the triggeron--pDM annihilation probability is close to 1. The most striking feature of the signal is its breadth: the spectrum extends to photon energies much lower than what one would typically expect from standard Hawking and final state radiation in PBHs. We expect that this is a generic feature of DS with pDM and mediator masses lighter than the PBH temperature resulting from the triggerons impart a large ``kick'' to the pDM, thereby broadening the final-state SM indirect detection signal.

The parameter values chosen for \Fig{fig:badm_spectra} give rise to a broad photon spectrum observable by the future AMEGO-X experiment. For $\anti{\chi}$ triggerons of mass similar to the peak energy of the Hawking distribution (\ie, $E_{\anti{\chi}} \approx E_{\rm pk}$), the typical $\anti{\chi} \chi$ annihilation cross section is
\beq
    \sigma_\ann \approx \frac{y^2}{64 \pi m_\chi^2} \approx 
    5 \times 10^{-3}~\MeV^{-2} \bl( \frac{y}{3} \br)^2 \bl( \frac{3~\MeV}{m_\chi} \br)^2 \ ,
\eeq
which corresponds to an annihilation coefficient of
\beq\label{eq:Lambda_badm}
    \Lambda = \frac{f_\chi \rho_0 \sigma_\ann r_s}{m_\chi} \approx 3.7 \, \bl( \frac{f_\chi}{0.2} \br) \bl( \frac{y}{3} \br)^2 \bl( \frac{3~\MeV}{m_\chi} \br)^3 \bl( \frac{\rho_0}{0.838~\GeV \cm^{-3}} \br) \bl( \frac{r_s}{11~\kpc} \br) \ ,
\eeq
where $f_\chi$ is the mass fraction of the pDM $\chi$, and we have used the Milky Way values of $\rho_0$ and $r_s$ as benchmarks \cite{deSalas:2019pee}. Note that the perturbative limit for the scalar quartic coupling is $y \approx 16 \pi^2$, well above the benchmark shown. The $f_\chi = 0.2$ reference value was chosen to avoid the Bullet Cluster's constraint on the fraction of DM that can have significant self-interactions \cite{Markevitch:2003at,Randall:2008ppe,Robertson:2016xjh}. Since these $\chi\chi$ self-interactions involve the coupling $\lambda$ of \Eq{eq:badm_lagrangian}, they can also be made significantly smaller by fine-tuning their tree-level ($\propto \lambda$) and loop-level contributions (a $\phi$-loop $\propto y^2$) \cite{Agashe:2020luo}. Doing so allows $f_\chi = 1$.

Finally, the $\phi$ ALP decays need to be quick enough for the local limit to hold. The ALP--photon coupling in \Eq{eq:badm_lagrangian} leads to $\phi \to \gamma\gamma$ decays, whose rate in the $\phi$ RF is given by
\bea
    \Gamma_\phi = \frac{g_{\phi\gamma\gamma}^2 m_\phi^3}{64 \pi} & \approx & 5 \times 10^{-34}~\GeV \ \bl( \frac{g_{\phi\gamma\gamma}}{10^{-11}~\GeV^{-1}} \br)^2 \bl( \frac{m_\phi}{1~\MeV} \br)^3 \nonumber\\
    & \approx & \bl( 40~\yr \br)^{-1} \ \bl( \frac{g_{\phi\gamma\gamma}}{10^{-11}~\GeV^{-1}} \br)^2 \bl( \frac{m_\phi}{1~\MeV} \br)^3 \ ,
\eea
where we have chosen benchmark values for $m_\phi$ and $g_{\phi\gamma\gamma}$ to be safe from robust astrophysical and cosmological constraints \cite{Depta:2020wmr,Dolan:2021rya,Balazs:2022tjl,Langhoff:2022bij,AxionLimits}. Since the ALPs come out of the $\anti{\chi}\chi$ annihilations boosted with a typical Lorentz factor of $E_{\rm pk} / m_\phi \sim \mathcal{O}(\text{few}) \times (T_\PBH / 1~\MeV) (1~\MeV / m_\phi)$, the ALP lifetime in the GF is still less than $1~\sec$. This means that the $\phi$s decay very close to their birthplace at the location of the $\anti{\chi}\chi$ annihilation in the galactic scale, and thus well before reaching the Earth.

\begin{figure}\centering
\includegraphics[width=0.75\linewidth]{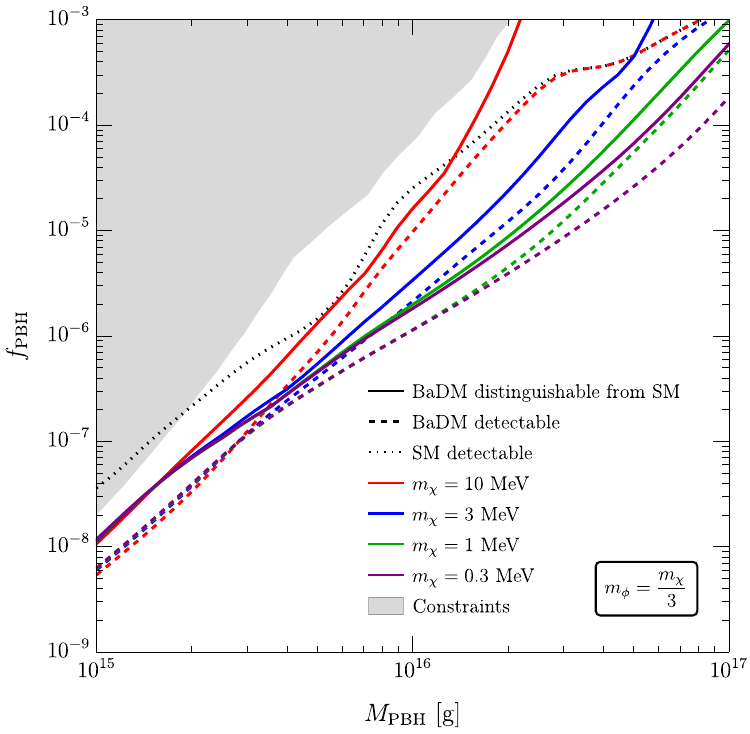}
\caption{Contours for detection and differentiation of the BaDM model with future AMEGO-X observation of GC gamma-rays. The dotted black line is the minimum $f_{\rm PBH}$ at a given $M_{\rm PBH}$ for which AMEGO-X can detect the corresponding PBH Hawking radiation spectrum above the background, if only SM particles are emitted. The dashed lines are the same as the dotted line, but for the BaDM model. The solid lines are the minimum requisite $f_{\rm PBH}$ for which we can distinguish the BaDM model from the SM, which means the spectrum would not be explained by the SM with a monochromatic PBH mass distribution. Different colors are for different mass choices, where red, blue, green, and purple correspond to $m_\chi=10~\MeV,~3~\MeV,~1~\MeV$, and $0.3~\MeV$, respectively. $m_\phi$ is chosen to be a third of $m_\chi$. The gray region is the constrained region from \cite{Clark:2018ghm,Coogan:2020tuf}. Note that colored lines are mostly below the dotted line since the peak of BaDM can be higher than the SM peak as shown in \Fig{fig:badm_spectra}. We computed the photon differential flux in the local limit ($\Lambda \gg 1$).}
\label{fig:BADMcontours}
\end{figure}

Let us now study the BaDM parameter space for which the triggered indirect detection signal could be observed by the AMEGO-X experiment, once again in the $\Lambda \gg 1$ local limit. In \Fig{fig:BADMcontours}, we show the $2\sigma$ AMEGO-X sensitivity \cite{Caputo:2022xpx} for detection of Hawking radiation, assuming the SM and BaDM cases, as well as for differentiating between them. The dotted line is the sensitivity for which Hawking radiation from the SM case can be detected in AMEGO-X, while the dashed lines are for detecting the BaDM model. The solid lines show the minimum value of $f_{\rm PBH}$ that allows AMEGO-X to distinguish between the BaDM model and the SM case. See \Sec{subsec:amegox} for the details. Different colors indicate different mass choices. Note that we assume the mass distribution of PBH is monochromatic. In principle, even if we measure two peaks in the Hawking radiation spectrum, these could come from the SM with a different PBH mass spectrum rather than from BaDM. However, many PBH production mechanisms predict a narrow PBH mass distribution, so it is unlikely to have two peaks in the SM case. Finally, note that in the SM+$\phi$ case, which was considered in \cite{Agashe:2022phd}, the spectrum is similar to the BaDM model, as shown in \Fig{fig:badm_spectra}. This means that it is very hard to distinguish between both scenarios. Thus, the solid lines in \Fig{fig:BADMcontours} should be interpreted not as the sensitivity curves to discover the labeled BaDM benchmarks, but instead as providing insights regarding where we can find new physics from Hawking radiation.

\subsection{Line-like spectrum: Dipole Transition DM (DTDM) model}\label{subsec:dtdm}

The second kind of signal we consider is a monochromatic line. Line signals can arise if the DS includes an excited dark state or heavy partner of the pDM,\footnote{We remind the reader that non-trivial particle spectra and their mutual interactions are ubiquitous in the SM, and they could very well be part of a DS as well, according to the principle of dark complexity.} which subsequently decays into photons.\footnote{Atomic dark matter presents an example of such ``excited partners'' of DM. Dark photons from the PBH could lift up the atomic DM into an excited state which will, however, decay back into dark photons. Decays into visible photons could occur via a kinetic mixing with the dark photons, but the signal would then be suppressed by this mixing. Furthermore, such a line signal would be ``buried'' under the PBH Hawking radiation, which has many more photons with the same energy.} Denoting the lighter, stable DM particle by $\chi_1$ and its heavier, unstable partner by $\chi_2$, the photon signal in question can arise from an inelastic dipole interaction of the form $d \, \anti{\chi}_2 \sigma^{\mu \nu} \chi_1 F_{\mu \nu}$ (for early work on dipole DM see Ref.~\cite{Sigurdson:2004zp}; for inelastic dipole transitions see Ref.~\cite{Profumo:2006im}). In the rest frame of $\chi_2$, the monochromatic photon has energy $E_{\gamma*}^{\rm [RF]} = \Delta m / \bl( 1 + \Delta m /(2 m) \br)$, where $\Delta m \equiv m_2 - m_1$ is the mass splitting between $\chi_2$ and $\chi_1$, and $m \equiv (m_1 + m_2) / 2$ is the average of their masses.

\begin{figure}
\centering
\includegraphics[width=0.7\linewidth]{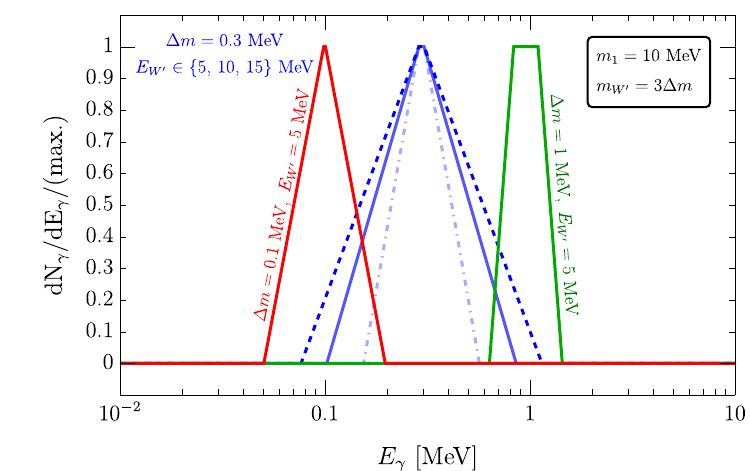}
\\
\includegraphics[width=0.7\linewidth]{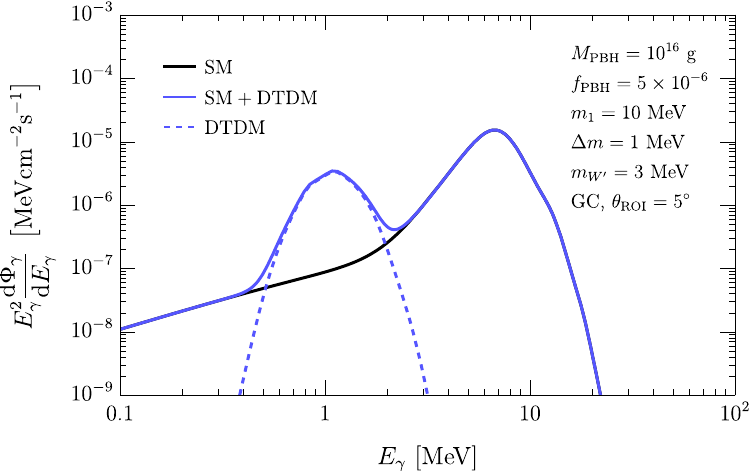}
\caption{{\it Top:} The trapezoid-shaped spectrum $dN_\gamma/dE_\gamma$ of \Eq{eq:dNgamma_general}, for the DTDM model, and various choices of the triggeron energy $E_\trg = E_{W'}$ and the mass difference $\Delta m$ between the DM $\chi_1$ and the mediator $\chi_2$. We have taken $m_1 = 10~\MeV$ and $m_{W'} = 3 \Delta m$. For convenience, we have normalized the spectra such that their maximum value is always $1$. The blue dot-dashed, solid, and dashed spectra correspond to $\Delta m = 0.3~\MeV$ but $E_{W'} = 5~\MeV$, $10~\MeV$, and $15~\MeV$, respectively. The red and green lines both have $E_{W'} = 5~\MeV$, but $\Delta m = 0.1~\MeV$ and $1~\MeV$, respectively. Note how a larger triggeron energy $E_{W'}$ results in more boosted photons, and thus a broader spectrum. Increasing $\Delta m$ shifts the peak of the spectrum (which is $E_{\gamma*} = \Delta m / (1 + \Delta m / (2m)) \approx \Delta m$) to larger values. It also narrows down the distribution, since larger $\Delta m$ means larger mediator mass $m_2 = m_1 + \Delta m$, and thus its RF has a smaller boost with respect to the GF for fixed triggeron energy $E_{W'}$. {\it Bottom:} The gamma-ray signal (photon spectral flux density times energy) in the DTDM model, coming from a ROI of $5\degree$ around the GC of the Milky Way. The black line is the expectation from the SM (\ie, direct Hawking radiation of photons plus final-state radiation contributions). The blue dashed line is the contribution to the signal from ambient $\chi_1$ pDM triggered by Hawking radiation by PBHs of $W'$ triggerons, following \Eq{eq:steps_dtdm}. The solid blue line is what the observer ultimately sees. For all curves, we have chosen the benchmark parameter values of $M_\PBH = 10^{16}~\gr$, $f_\PBH = 5 \times 10^{-6}$ ($1/100$ of the current bound for this mass \cite{Carr:2020gox}), $m_1 = 10~\MeV$, $\Delta m = 1~\MeV$, and $m_{W'} = 3~\MeV$. We computed the photon differential flux in the local limit ($\Lambda \gg 1$).}
\label{fig:dtdm_spectra}
\end{figure}

The indirect detection signal thus relies on having the mediator $\chi_2$ present, which is challenging to produce from SM particles at late times, when the temperature is significantly below the mass threshold. However, dark sector triggerons emitted by PBHs can convert the $\chi_1$s in the galactic halo into $\chi_2$s. As we show in more detail in \App{app:C}, a spontaneously broken $\SU{2}'$ dark gauge symmetry can realize this scenario, with the pDM $\chi_1$ and its heavier partner $\chi_2$ being a linear combination of the lower and upper components of a $\SU{2}'$ doublet, and the massive dark $W^{a \prime}$ gauge bosons playing the role of the triggerons.\footnote{In general all three dark gauge bosons participate as triggerons. However, in the limit where $\chi_{1,2}$ are identical to the lower and upper entries of the $\SU{2}'$ doublet, the triggeron is identified with $W^{\prime +} \equiv \bl( W^{\prime 1} - i W^{\prime 2} \br)/\sqrt{2}$; see \App{app:C} for details.} In summary, the steps of \Eq{eq:steps} in this {\it ``dipole transition dark matter''} (DTDM) model are
\bea\label{eq:steps_dtdm}
    \text{DTDM model:} \quad 
    \begin{cases}
        \text{emission: } \quad & \text{PBH } \to W^{a \prime} \ ; \\
        \text{annihilations: } & W^{a \prime} \chi_1 \to W^{b \prime} \chi_2 \ ; \\
        \text{decays: }  \quad & \chi_2 \to \chi_1 \gamma \ ; \\
        \text{(but not: } \quad & \chi_2 \, \cancel{\to} \, \chi_1 W') \ ,
    \end{cases}
\eea
with the kinematic requirements necessary for a line-like $\gamma$-ray signal being
\beq
    \Delta m < m_{W'} \lesssim \mathcal{O}(\text{few}) \times T_\PBH \ll m \ . \label{eq:kin_dtdm}
\eeq
The inelastic scattering $W^{a\prime} \chi_1 \to W^{b\prime} \chi_2$ can occur since the kinematic conditions of \Eqs{eq:kin1}{eq:kin2} are satisfied. In particular, the triggering $W^{a \prime}$ is energetic enough to overcome the mass splitting $\Delta m$, \ie, $E_\trg - m_\trg = E_{W'} - m_{W'} > \Delta m$. Moreover, we assume $m_{W'} > \Delta m$, in order to prevent $\chi_2$ from decaying back to $\chi_1$ so that it can instead decay to the SM photon. In order for the photon signal to be a line that is easier to observe, the $\chi_2$ particle must be produced basically at rest. This means that the pDM $\chi_1$ must be sufficiently heavy so that the ``kick'' it receives from the triggeron does not result in a large recoil, \ie,  $m_{1,2} \gg E_{W'} \sim T_\PBH$. Ignoring the subdominant DM halo velocity, the width of the photon line is then typically $\sim T_\PBH / m \ll 1$. Thus, the RF monochromatic line signal is only mildly broadened into a trapezoidal spectrum in the GF, retaining its line-like appearance centrered at $E_{\gamma*} = \Delta m/(1 + \Delta m / (2m)) \approx \Delta m$. The upper panel of \Fig{fig:dtdm_spectra} shows $d N_\gamma / d E_\gamma$ for some values of $m$, $\Delta m$, $m_{W'}$, and $E_\trg = E_{W'}$.

The appeal of this scenario is evident: A hypothetical dark sector indirect detection signal, like the one described here, can have an energy $E_{\rm signal} \sim \Delta m$ that is {\it a priori} different from the typical energy $E_\trg \sim T_\PBH$ of the PBH Hawking radiation that triggered the signal. A photon line stands in sharp contrast to the broad spectrum of Hawking radiation; the fact that $m \gg T_\PBH$ means that no $\chi_2$s are emitted by the PBHs, and thus the entirety of the signal arises from the triggeron interactions with ambient pDM. In other words, the photon signals coming both from rich DS physics and the Hawking radiation of PBHs will be correlated in their amplitudes (as described in the previous section) but clearly distinguishable in both their energies and spectra. The lower panel of \Fig{fig:dtdm_spectra} shows the indirect detection photon signal associated with our DTDM model, for parameter values that give rise to a signal detectable by AMEGO-X, and in the local limit ($\Lambda \gg 1$) of the signal, where the triggeron--pDM annihilation probability is set to 1.

Leaving the model-building details to \App{app:C}, here we give some benchmark values for the various quantities governing the indirect detection signal.\footnote{There are several other details that merit careful study beyond the scope of this paper. For example, in the minimal version of the DTDM model considered here, the massive $W^{a\prime}$ bosons are stable, and thus contribute to the DM density, although in a subdominant role due to their smaller mass. In addition, an asymmetry-generation mechanism must be present in the early Universe so that the $\chi$ particles constitute the dominant pDM component, thereby avoiding $\chi_1 \anti{\chi}_1 \to W' W'$ annihilations. Since the purpose of this section and of \App{app:C} is simply to provide a plausible particle physics origin for the signal template under study, we leave these considerations of the thermal history of the model to future work.} For example, the $W^{a \prime} \chi_1 \to W^{b \prime} \chi_2$ annihilation cross section is estimated as
\beq\label{eq:sigma_chiW}
    \sigma_\ann \approx \frac{g^{\prime 4} \abs{V}^4}{16 \pi m_1^2}
    \approx 0.02~\MeV^{-2} \bl( \frac{g' \abs{V}}{3} \br)^4 \bl( \frac{10~\MeV}{m_1} \br)^2 \ ,
\eeq
where $g'$ is the dark gauge coupling, and $\abs{V}$ schematically represents the appropriate mixing factors that relate the interacting basis of the $\chi$ fermions to the mass eigenstates $\chi_{1,2}$. The corresponding annihilation coefficient is
\beq
    \Lambda = \frac{f_1 \rho_0 \sigma_\ann r_s}{m_1} \approx 3.6 \, \bl( \frac{f_1}{0.2} \br) \bl( \frac{g' \abs{V}}{3} \br)^4 \bl( \frac{10~\MeV}{m_1} \br)^3 \bl( \frac{\rho_0}{0.838~\GeV \cm^{-3}} \br) \bl(  \frac{r_s}{11~\kpc}\br) \ , \label{eq:Lambda_dtdm}
\eeq
where $f_1 \equiv \rho_1 / \rho_0$ is the mass fraction of the pDM $\chi_1$. The $\chi_1 \chi_1$ self-interaction cross section is very large, $\sigma_{\rm self} \sim 4 \times 10^5~\cm^2 / \gr \ \bl( g/3 \br)^4 \bl( m_1 / 10~\MeV \br)^2 \bl( 1~\MeV / m_{W'} \br)^4$ (since the typical momentum exchanged in the halo is $m_1 v_{\rm halo} \sim 10~\keV \bl( m_1 / 10~\MeV \br) \ll m_{W'} \sim \mathcal{O}(1~\MeV)$), so we have chosen a Bullet Cluster-safe mass fraction $f_1$.\footnote{Dark disk formation cannot happen, since the typical $\chi_1$ pDM kinetic energy is much smaller than the mass of the dark gauge bosons $W'$, and thus the $\chi_1$s cannot emit $W'$s to cool the halo.}

A typical dipole value in our model is $d \approx 10^{-8}~\GeV^{-1} \approx 6 \times 10^{-19}~e \, \cm$ (see \App{app:C} for details), which is safe from Lyman-$\alpha$ and SN 1987A constraints \cite{Profumo:2006im}, and well below $2 \times 10^{-3}~\GeV^{-1} \approx 10^{-16}~e \, \cm$, the scale of the LEP 90\% C. L. bounds on the magnetic moment of light dark particles \cite{Fortin:2011hv}. Such a dipole gives $\chi_2$ a RF decay rate $\Gamma_2$ given by
\bea
    \Gamma_2 \simeq \frac{d^2 \Delta m^3}{8 \pi} & \approx & 4 \times 10^{-30}~\GeV \bl( \frac{d}{10^{-8}~
    \GeV^{-1}} \br)^2 \bl( \frac{\Delta m}{0.1~\MeV} \br)^3\,, \nonumber\\
    & \approx & \bl( 2 \times 10^5 \sec \br)^{-1} \bl( \frac{d}{10^{-8}~
    \GeV^{-1}} \br)^2 \bl( \frac{\Delta m}{0.1~\MeV} \br)^3 \ , \label{eq:Gamma2}
\eea
which is fast enough for the $\chi_2$ particles to decay well within a parsec of being created. Considering also the non-relativistic $\chi_2$ recoil velocities, the decay happens well before the particle reaches the Earth.

\begin{figure}\centering
\includegraphics[width=0.75\linewidth]{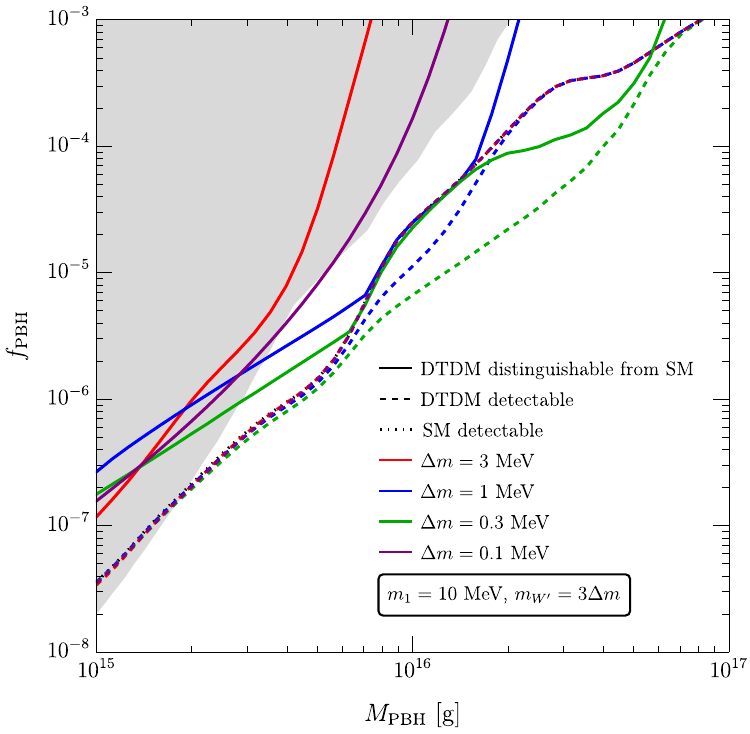}
\caption{Contours for detection and differentiation of the DTDM model with future AMEGO-X observation of GC gamma-rays. The dotted black line (mostly overlapped with the red and purple dashed lines) is the minimum $f_{\rm PBH}$ at a given $M_{\rm PBH}$  for which AMEGO-X can detect the corresponding PBH Hawking radiation spectrum above the background, if only SM particles are emitted. The dashed lines are the same as the dotted line, but for the DTDM model. The solid lines are the minimum requisite $f_{\rm PBH}$ for which we can distinguish the DTDM model from the SM, which means the spectrum would not be explained by SM spectrum with delta-function mass distribution. Different colors are for different mass choices, where red, blue, green, and purple correspond to $\Delta m = 3~\MeV,~1~\MeV,~0.3~\MeV$, and $0.1~\MeV$. We choose $m_1=10~\MeV,~m_{W'}=3\Delta m$, and $\Lambda\to\infty$ here. The gray region is the constrained region from \cite{Clark:2018ghm,Coogan:2020tuf}. Note that the order of colored lines in the figure is different from the mass ordering as AMEGO-X sensitivity is not uniform over the photon energies. We computed the photon differential flux in the local limit ($\Lambda \gg 1$).}
\label{fig:DTDMcontours}
\end{figure}

We are now in a position to work out the indirect detection photon signal of the DTDM model, and study the parameters for which it could be observed by the AMEGO-X experiment, in the $\Lambda \gg 1$ local limit. 
In \Fig{fig:DTDMcontours}, we show the $2\sigma$ AMEGO-X sensitivity \cite{Caputo:2022xpx} for detection of the Hawking radiation assuming the SM and DTDM cases, as well as for differentiating between them. In a similar way to \Fig{fig:BADMcontours}, the dotted line denotes the Hawking radiation detection threshold for the SM case, the dashed lines correspond to the the DTDM case, and the solid lines to when the spectra from each can be distinguished. Different colors indicate different parameter choices.
The location of the DTDM peak in the photon spectrum depends on $\Delta m$, as shown in \Fig{fig:dtdm_spectra}. Note that it is hard to distinguish the DTDM model from the SM case if $\Delta m$ is outside of the AMEGO-X sensitivity range. As a result, the red ($\Delta m = 3~\MeV$) and purple ($\Delta m = 0.1~\MeV$) detection contours (dashed lines) are almost identical to the SM detection line, while the differentiation contours (solid lines) are much above the dashed lines, requiring larger values of $f_\PBH$. On the other hand, the blue ($\Delta m = 1~\MeV$) and the green ($\Delta m = 0.3~\MeV$) lines, for which $\Delta m$ is near the energy for which AMEGO-X has the most sensitivity, have lower detection and differentiation sensitivities.

\section{Discussion and Conclusion}
\label{sec:conclusions}

In this paper we studied the possibility of observing indirect detection signals of dark sector processes. If dark matter (DM) is part of a broader dark sector (DS) with a rich phenomenology (what has been called {\it dark complexity}), dark processes may in principle give rise to a standard model (SM) signal through a mediator. Despite the rich dark physics, the DM remains ``quiet'' if the DS processes required to produce a mediator and an eventual SM signal involve dark particles not readily available. Since these ``{\it triggerons}'' are dark, they may not be produced through SM means. They can, however, be emitted as Hawking radiation from black holes (BHs), if their mass is sufficiently light compared to the Hawking temperature. Since the emission of Hawking radiation is ultimately a gravitational effect, BH-sourced triggerons will be readily available to activate DS phenomenology, as long as BHs are present in sufficient numbers.

Primordial black holes (PBHs) can be a form of DM, and those with masses between $5 \times 10^{14}~\gr$ and $\sim 10^{17}~\gr$ can be hot enough and abundant enough to produce a steady source of triggerons, thereby inducing an indirect detection signal on the ambient DM. In other words, the PBHs can act as catalysts of otherwise inaccessible DS physics. As we have shown, the PBHs need not represent a significant fraction of the DM for this signal to be observable. The phenomenology of PBHs and particle DM has been extensively anlyzed in the past, but mostly in isolation. In this work we have initiated the study of new signals emanating from a non-trivial interplay between the simultaneous presence of these two different kinds of DM.

As a proof of principle, we considered two DS models where a particle DM (pDM) indirect detection signal requires the activation of DS physics through PBH-sourced triggerons. Our first example, which we call the ``boosted anti-dark matter'' (BaDM) model, relies on an asymmetric DM (ADM) scenario. There, a lack of ambient DM antiparticles renders the DM particles around us effectively inert. However, the PBH population can emit the DM antiparticles, which then trigger the DS physics by annihilating with the ambient pDM and producing a lighter ALP mediator. The boosted mediator subsequently decays into photons, which can be observed as a broad spectrum. In our second model, which we called the ``dipole transition dark matter'' (DTDM), the pDM has an excited dark state ``partner'' to which it couples via an inelastic dipole coupling to the SM photon. Triggerons emitted by PBHs can then excite the ambient DM particle into its heavier partner, which subsequently decays into a photon line signal. As detailed in \App{app:C}, the DTDM scenario can be realized if the pDM is a linear combination of the up and down components of a dark $\SU{2}'$ doublet, with its heavier partner being the orthogonal combination. The triggerons are the $\SU{2}'$ gauge bosons (which gain a mass upon spontaneous symmetry breaking), and their interactions with the pDM generate the heavier partner. $\TeV$-mass particles charged under both $\SU{2}'$ and QED can induce the required dipole.

We want to emphasize that, although we have considered two specific DS models to illustrate our point, many other scenarios also have triggerons that can lead to similar signals.  For example, in models where the constituents of a composite DM are (milli)charged, transitions between excited and ground states could lead to detectable absorption or emission photon signals ~\cite{Profumo:2006im,Kaplan:2009de,Kribs:2009fy,Cline:2012bz,Frandsen:2014lfa,Cline:2014eaa,Ganguly:2023rzm}. Such signals could also be triggered by either dark or visible Hawking radiation, depending on the properties of the DM. Alternatively, dark axion portal models could also generate a line signal (for early work see Refs.~\cite{Kaneta:2016wvf,Kaneta:2017wfh}; for a model where this coupling is the dominant one see Ref.~\cite{Hook:2023smg}; for recent constraints see Ref.~\cite{Hong:2023fcy}). Indeed, if the pDM is made of an axion $a$ or dark photon $A'$, the $g_{a \gamma \gamma'} a F \tilde{F}'$ coupling present in these models can lead to observable DM decays. In the presence of PBH emission of $A'$ ($a$) radiation, the DM particles $a$ ($A'$) can undergo stimulated decays, thereby producing an enhanced photon signal. Yet another class of models giving qualitatively different spectra is that in which the DS portal couples to SM particles other than the photon, such as charged leptons or neutrinos.
Lastly, PBHs can produce bosonic triggerons via the superradiance process (in analogy to ALPs in Refs.~\cite{Dent:2024yje, Branco:2023frw}), with triggerons being gravitationally bound in a superradiance cloud around the PBH and subsequently annihilating with ambient pDM in the halo regions swept by the cloud. This scenario exhibits a distinct morphological feature: triggered signals are proportional to $\rho^2_\DM$, compared to the $\rho_\DM$ dependence of primary Hawking radiation.

Returning to the concrete BaDM and DTDM models, in this paper we have shown that the triggered indirect detection signals, which are strongest in the $\sim \MeV$ range, can be observed by future gamma-ray experiments such as AMEGO-X \cite{Caputo:2022xpx} and e-ASTROGAM \cite{e-ASTROGAM:2016bph}.
We showed the normalized photon spectrum from a single triggeron (top panels) and the photon flux expected at Earth (bottom panels) for various parameter choices of the BaDM model in \Fig{fig:badm_spectra} and of the DTDM model in \Fig{fig:dtdm_spectra}. The peak location of the photon flux from the BaDM model does not depend much on its parameters, while the mass splitting $\Delta m$ between the DM and its heavier partner clearly determines the peak location for the DTDM model.
Especially for the DTDM model, this peak can be completely separated from the primary photon peak of the Hawking spectrum, resulting in a unique signal.
These additional peaks make the Hawking radiation spectrum distinguishable from the SM case; in \Sec{sec:models} we showed the sensitivity lines for the detection of these models as well as those indicating when we can distinguish them from the SM, \Fig{fig:BADMcontours} for BaDM, and \Fig{fig:DTDMcontours} for DTDM.

Our results capture the key features of PBH-triggered dark sector signals. Let us briefly discuss the uncertainties associated with this search strategy, as well as outline its potential future refinements. Currently, our signal flux calculation relies on the local limit of triggeron annihilations. Detailed methods for calculating the photon flux are explained in \App{app:B}. More accurate spectra could be obtained by accounting for {\em finite} values of the annihilation coefficient $\Lambda$ that reflect the ultimately non-local origin of the signal, incorporating the triggeron energy dependence of $\Lambda$ in the gamma-ray spectrum, and employing a more precisely calibrated halo model, which will necessarily require a more accurate modeling of the angular dependence of the signal. In addition, measuring a nearly SM-only Hawking radiation spectrum is feasible from targets with suppressed triggeron annihilation probability, such as dwarf galaxy halos. This can assist with modeling the null hypothesis, making it easier to identify dark sector signals. On the theoretical side, the precision of the Hawking radiation spectrum depends on accurately determining graybody factors, which could be important when considering the additional PBH spin parameters and different dark sector particle spins. Lastly, multi-messenger signals can offer useful information about PBH parameters. For example, gravitational waves from PBH formation processes have been found to be correlated with the PBH mass distribution and gamma-ray signals~\cite{Agashe:2022jgk}.

Let us reiterate that the specific BaDM and DTDM scenarios we considered are meant to illustrate the general mechanism that is the ``coupling'' between PBHs and ambient particle DM. These are toy models, particle physics realizations that exemplify what we expect are more general features of indirect detection signals of dark complexity in DS. The first lesson we have learned is that the largest possible indirect detection signals coming from triggered DS processes require $\sim \MeV$ pDM masses and large DS couplings couplings. This is because the activation of DS physics necessarily depends on the triggeron--pDM annihilation cross section and the pDM number density. In order to have a large signal, a triggeron mean-free-path shorter than the typical galactic length scale is required (what we called the ``local limit''), which results in the stated $\MeV$ pDM mass scales and $\mathcal{O}(1)$ couplings. Larger pDM masses and smaller couplings are of course possible, but lead to weaker signals. Of course, the main premise of our paper is that {\it if} pDM is part of a DS {\it and} it has a preference to interact with dark rather than SM particles, then indirect signals of this kind could be detected, so the required large couplings should not come as a surprise. The second lesson is that photon signals in the $\MeV$ range arise naturally in our scenario.
The main reason is the PBHs, which source the triggerons required for the signal to exist in the first place. The larger the triggeron Hawking radiation emission rate of the PBHs, the larger the signal. However, this also increases the evaporation rate of the PBHs. The PBHs with the largest emission rates which are yet to evaporate away entirely have masses above $5 \times 10^{14}~\gr$ and temperatures below $\sim 20~\MeV$. Slightly heavier PBHs can be abundant enough to produce triggerons in sufficient numbers. These triggerons have energies comparable to the PBH temperatures, namely $\lesssim \mathcal{O}(10~\MeV)$, which means the total energy available in the triggeron--pDM collisions, given the $\sim\MeV$ pDM masses mentioned above, is also $\MeV$. Thus, as exemplified in the BaDM scenario, the photon spectrum is also in the $\MeV$ range, and relatively broad (due to the boosts of the particles involved). This means that the indirect detection signal may be superimposed on the photon signal from direct PBH emission, and some effort must be put in in order to disentangle them. Nevertheless other signals, qualitatively different from the blackbody-like radiation of PBH photons, can also arise in sufficiently complex DS. Indeed, as demonstrated by the DTDM model, sharper features such as spectral lines at lower energies can be produced in scenarios with dark ``excited states'', although AMEGO-X will be unable to detect photons below $100~\keV$. Considering this situation, X-ray measurements can be further combined with gamma-ray data to cover a broader DS parameter space featuring soft transition lines. Broadly speaking, $\sim \MeV$ is the natural scale for PBH-induced DS signals, from independent considerations regarding the triggeron mean-free-path and the properties of the PBHs.

In summary, it is clear that we need to cast as wide a net as possible in order to catch DM signals. As we have shown, gamma-ray experiments searching for the Hawking radiation of PBHs may stumble upon unexpected surprises coming from a dark sector phenomenology hitherto inaccessible to the SM, but brought in contact by PBHs themselves. We hope that, by motivating the synergy between the physics of primordial black holes and ambient dark matter, we have also demonstrated that the future of darkness is bright.

\acknowledgments{

The authors thank Regina Caputo, Valerio De Luca, William DeRocco, Henrike Fleischhack, and Shmuel Nussinov for useful discussions. MBA thanks Stephanie Buen Abad for proofreading this manuscript. KA and MBA are supported by NSF grant PHY-2210361 and the Maryland Center for Fundamental Physics. JHC is supported by Fermi Research Alliance, LLC under Contract DE-AC02-07CH11359 with the U.S. Department of Energy. The research of YT is supported by the National Science Foundation Grant Number PHY-2112540 and PHY-2412701. YT would like to thank the Tom
and Carolyn Marquez Chair fund for its generous support. YT would also like to thank the
Aspen Center for Physics (supported by NSF grant PHY-2210452).
 BD is supported  by the U.S.~Department of Energy Grant DE-SC0010813. TX is supported by the U.S. Department of Energy under grant DE-SC0009956.}

\appendix

\section{Halo Evaporation and Triggeron Annihilations in DM Mini-halos}\label{app:A}

In this appendix, we discuss two potential issues with our scenario. The first is the question of whether the triggeron--pDM annihilations can lead to significant evaporation of the DM halo. The second is whether the triggerons will annihilate in what has been variously called in the literature the PBH's DM ``mini-halo'' or ``spike'', a population of accreted (particle) DM in the immediate vicinity of the PBH and gravitationally bound to it \cite{Mack:2006gz,Ricotti:2007jk,Ricotti:2007au,Ricotti:2009bs}. As we demonstrate, the answers to both questions are in the negative for the parameter space of interest.

\subsection{Halo Evaporation}\label{app:A1}

Our first concern should be the impact that triggeron--pDM interactions have on the DM halo. In particular, the annihilations in the second step of \Eq{eq:steps_general} may result in forbiddingly large halo evaporation. To estimate the halo mass loss due to pDM triggering, we can assume for simplicity that the triggeron production rate is given by the number of PBHs in the galactic halo multiplied by the PBH evaporation rate, which can be written in terms of the initial halo mass $M_0$ and the PBH's mass fraction $f_\PBH$ as $\Gamma_\PBH N_\PBH = \Gamma_\PBH f_\PBH M_0/M_\PBH$. If a triggeron has a total probability $\mathcal{P}_\ann$ of annihilating with a pDM in the halo, the change in the pDM's number is given by $\dot N_\pDM = - \mathcal{P}_\ann \Gamma_\PBH N_\PBH$. Finally, by conservatively assuming that the annihilation products are not gravitationally bound to the galaxy, we can relate the halo mass loss $\dot M_{\rm halo}$ to the change in the number of DM particles (ignoring the loss due to decreasing evaporating PBH mass):
\beq
    \dot{M}_{\rm halo} = m_\pDM \dot{N}_\pDM = - \frac{m_\pDM}{M_\PBH} \mathcal{P}_\ann \Gamma_\PBH f_\PBH M_0 \ .
\eeq

In general $\mathcal{P}_\ann$ is a function of the halo and the pDM properties, such as the halo size, triggeron--pDM annihilation cross section, and the pDM abundance (which is itself time-dependent). For our current purposes, it suffices to conservatively take $\mathcal{P}_\ann = 1$. In that case, the mass fraction $f_\pDM = m_\pDM N_\pDM / M_0$ of the pDM with respect to the initial halo mass evolves as
\beq
    f_\pDM (t) \simeq f_{\pDM, 0} - \frac{m_\pDM}{M_\PBH} f_\PBH \Gamma_\PBH \, t
\eeq
where $t$ is the age of the galaxy. Taking into account the existing bounds on the PBH mass fraction \cite{Carr:2020gox} and a typical galactic age of $\sim 10~\Gyr$, we find that the amount of evaporation that has taken place is very small:
\begin{widetext}
\bea
    \Delta M_{\rm halo} = M_0 - M(t)
    & \approx & 2.5 \times 10^{-9} \, M_0 \ \bl( \frac{m_\pDM}{10~\MeV} \br) \bl( \frac{10^{15}~\gr}{M_\PBH} \br)^2 \bl( \frac{f_\PBH(M_\PBH)}{1.5 \times 10^{-7}} \br) \bl( \frac{t}{10~\Gyr} \br) \nonumber\\
    & = & 2.5 \times 10^3 \, M_\odot \, \bl( \frac{M_0}{10^{12} M_\odot} \br) \ \bl( \frac{m_\pDM}{10~\MeV} \br) \bl( \frac{10^{15}~\gr}{M_\PBH} \br)^2 \bl( \frac{f_\PBH(M_\PBH)}{1.5 \times 10^{-7}} \br) \bl( \frac{t}{10~\Gyr} \br) \ .
\eea
\end{widetext}
Therefore, we can simply assume that the total halo mass (and that of its PBH and pDM constituents) remains constant and equal to $M_0$ throughout the timescales of interest.

\subsection{Triggeron Annihilation in DM Spike}\label{app:A2}

Since PBHs are formed in the very early Universe, they serve as seeds for the accretion of pDM. As a result, the pDM density near PBHs today can be significantly larger than the average local density \cite{Mack:2006gz,Ricotti:2007jk,Ricotti:2007au,Ricotti:2009bs}. The triggerons emitted by PBH Hawking radiation can therefore annihilate in this pDM ``spike'', which may increase $\Lambda$ as defined in \Eq{eq:Lambda}. However, the typical size of the spike is much smaller than the typical galactic distance ($\sim r_s$), leading to a huge suppression of the probability for triggeron--pDM annihilations to take place inside the spike, compared to them occurring in the galactic halo instead. For example, taking the simulation results found in Ref.~\cite{Boudaud:2021irr} as a benchmark, a central density of $1~{\rm GeV}$ pDM in the spike of a $10^{-18}M_{\odot}$ ($\approx 2 \times 10^{15}~\gr$) PBH is roughly $10^4~\gr \, \cm^{-3}$, with a density profile scaling as $\rho\propto r^{-3/2}$ in the outer region of the spike. The annihilation probability is proportional to the line-of-sight integral of the pDM density inside the spike. Thus, we have checked that the spike contribution to $\Lambda$ is about a factor of $10^{4}$ smaller than that of the Milky Way halo. Smaller pDM masses result in an even more severe suppression. Since $\Lambda$ in the galaxy is already only an $\mathcal{O}(1)$ number for our benchmark parameters we conclude that, even under the conservative assumption of a power-law index of $3/2$, the contribution to $\Lambda$ from pDM spikes can be neglected. Therefore we consider only the galactic contribution to $\Lambda$ throughout this paper, ignoring the pDM spikes of PBHs.

\section{Details of the Differential Photon Flux}\label{app:B}

\begin{figure}
\centering
\includegraphics[width=0.5\linewidth]{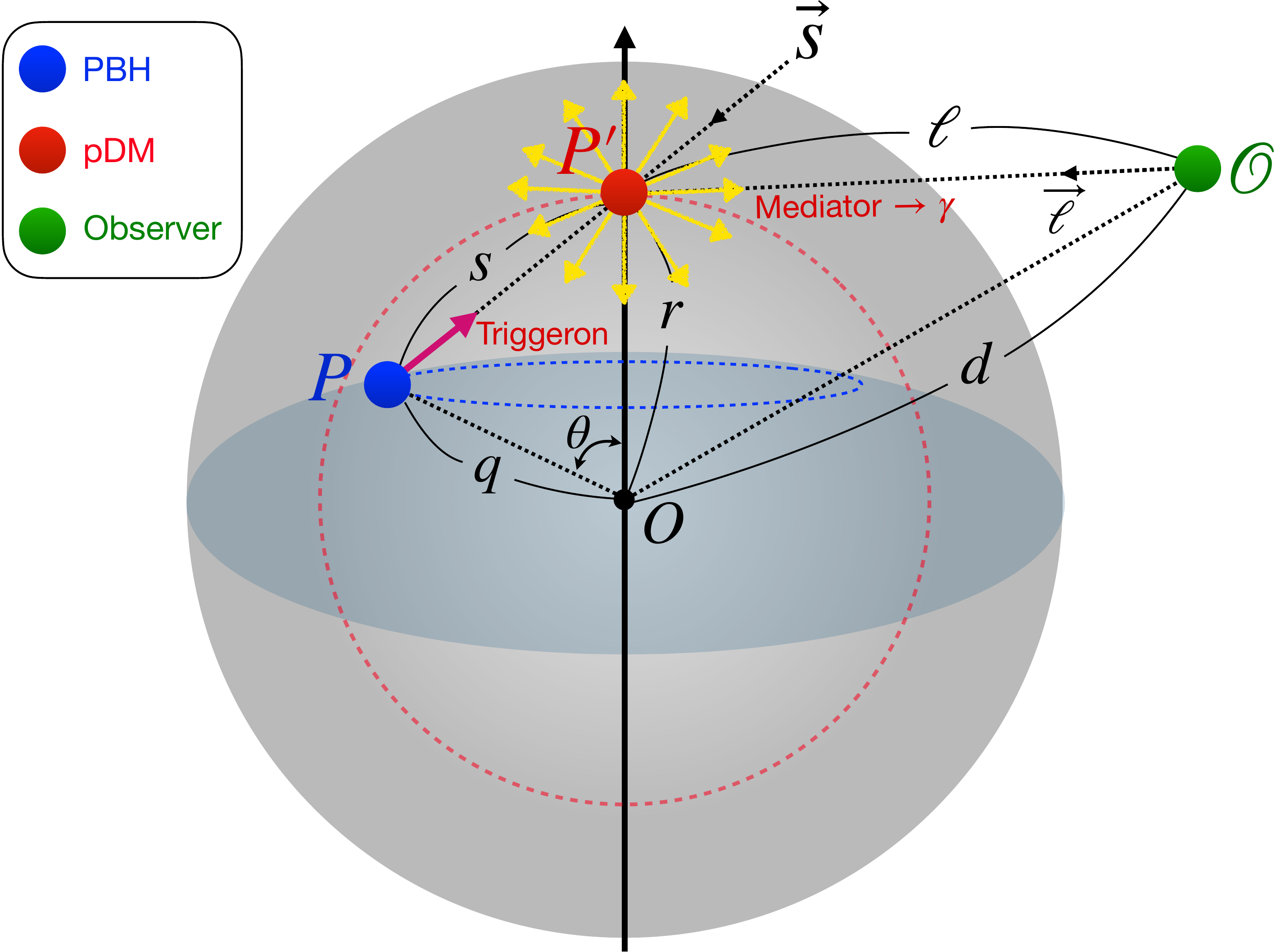}
\caption{The geometry of PBH-induced dark sector indirect detection signals. The gray sphere represents the DM halo, with the vertical direction corresponding to $\hat{\mathbf{k}}$ in the text. The PBHs located at $P$ (blue dot) isotropically emit triggerons via Hawking radiation. These travel a distance $s$ before annihilating with ambient pDM at the location $P'$ (red dot), a distance $r$ from the galactic center (GC) $O$ (black dot). The point $P$ is at a distance $q$ from the GC. The DM--SM mediators produced in these annihilations promptly decay into photons at a location $P'' = P'$, producing a gamma-ray signal. The observer is represented by $\mathcal{O}$ (green dot), and is located at a distance $d$ from $O$, and with a line-of-sight distance $\ell$ to the location $P'$ at which the photons are emitted. While in this picture we have placed $\mathcal{O}$ outside the DM halo (as it is be the case should the ROI correspond to a dwarf galaxy), this does not have to be the case. Due to the evident axial symmetry of the geometry, all the points in the blue dotted circle contribute equally to the signal at $P'$. If $\mathcal{O}$ is very far away from all the points in the ROI, $\ell \approx d$ and all points in the dotted red sphere contribute equally to the signal.}
\label{fig:geometry}
\end{figure}

We devote this appendix to the study of the differential photon flux signal $d \Phi_\gamma / d E_\gamma$ produced by the triggering of pDM by PBHs. After a thorough description of its derivation, we obtain useful analytic limits of this flux under various simplifying assumptions, in particular obtaining \Eq{eq:dPhidE_simplest}. We utilize the methodology and notation of Ref.~\cite{Agashe:2020luo}, with some minor variations.
On the subject of notation, throughout this appendix we work in three distinct frames of reference: the galactic frame (GF), the center-of-mass frame (CF) of the triggeron--pDM annihilation event, and the mediator's rest frame (RF). For those kinematic variables that take different values in different frames, we specify the frame of reference in question as a superindex in square brackets (\ie, [XF] for the ``X frame''). Whenever this index is omitted, the kinematic variable is evaluated in the galactic frame which, since we neglect the Sun's and Earth's motion in this paper, is the same as the observer's frame. For example, $E_\gamma^{\rm [RF]}$ is the photon energy in the rest frame of the mediator, while $E_\gamma$ is the same photon's energy but in the observer/galactic frame. We encourage the reader to follow along our discussion below with the help of \Fig{fig:geometry}.

We begin by assuming that the DM, \ie~both in its PBH and pDM populations, is distributed in a spherically-symmetric halo, represented by the gray sphere in \Fig{fig:geometry}. At every point $P'$ (red dot in \Fig{fig:geometry}) in the halo the pDM is shined upon by the incoming triggeron flux emitted by the PBHs at all other points $P$ (blue dot).\footnote{Since the number densities of both asteroid-mass PBHs and pDM are extremely large.} The result of these triggeron--pDM annihilations is the production of a mediator flux which, after some time, decays into a photon signal (yellow arrows). Our goal is to understand this final photon flux as seen by an observer at a point $\mathcal{O}$ (green dot) a distance $d$ from the galactic center (GC) $O$ (black dot). Taking $O$ as the origin of our galactic coordinate system, let us consider the pDM annihilations and subsequent mediator production taking place at a differential element of volume $dV_{P'}$ located at $P'$ and with position vector $\mathbf{r}$. Let $dV_P$ denote another differential element of volume centered at $P$; let $\mathbf{s}$ denote distance vector $P'P$, and let $\mathbf{q}(\mathbf{r}, \mathbf{s}) \equiv \mathbf{r} + \mathbf{s}$ denote the position vector of $P$. If the mediator's lifetime is shorter than the typical (\ie, kiloparsec) galactic scales, as is the case for the two toy models considered in this work, the point $P''$ at which the mediator decays and gives rise to the photon signal in question is very close to $P'$. Henceforth we work under this assumption and take $P'' = P'$.

The fundamental quantity that will eventually give us $d \Phi_\gamma / d E_\gamma$ at $\mathcal{O}$ is the angular and energy distribution of the rate at which photons are created by the annihilation at $P'$ of the triggerons emitted at $P$, as a function of the triggeron energy $E_\trg$. This verbiage can be written more succinctly as the following quantity:
\beq\label{eq:d6N}
    \frac{d^6 N_\gamma}{dE_\gamma d \Omega_{\gamma P'} dt dE_\trg dV_{P'} dV_P} = \frac{d^4 N_\ann}{dt dE_\trg dV_{P'} dV_P} \ \frac{d^2 N_\gamma}{dE_\gamma d\Omega_{\gamma P'}} \ .
\eeq
The first factor, $d^4 N_\gamma / dt dE_\trg dV_{P'} dV_P$, is the triggeron--pDM annihilation rate per unit volume at $P'$ per unit triggeron energy, caused by the triggerons produced at $P$. The second factor, $d^2 N_\gamma / dE_\gamma d\Omega_{\gamma P'}$, is the angular and energy distribution of the photons created at $P'$ from the decays of the mediators produced in these annihilations. Here $\Omega_{\gamma P'}$ denotes the angular components of the photons' momenta, \ie~the coordinates of the celestial sphere $S^2_{P'}$ centered at $P'$, and {\it not} the angular components of $\mathbf{r}$ (which are part of $d V_{P'}$). This distribution is normalized such that $\int\dd E_\gamma \dd \Omega_{\gamma P'} \, d^2 N_\gamma / dE_\gamma d\Omega_{\gamma P'}$ equals the number of photons produced in each annihilation event. For example, there are 4 photons per annihilation in the BaDM model, while only 1 per annihilation in the DTDM model.

The differential flux at $\mathcal{O}$ along a line-of-sight (los) distance $\ell$ to a region-of-interest (ROI) of solid angle $\Omega_{\rm ROI}$ is given by
\beq\label{eq:dPhidE}
    \frac{d \Phi_\gamma}{dE_\gamma} = \frac{1}{4\pi} \int\limits_{\rm ROI} \dd \Omega \, \int\limits_{\rm los} \dd \ell \, \bl[ \int\dd E_\trg \int\dd V_P \ \bl( 4 \pi \, \frac{d^6 N_\gamma}{dE_\gamma d\Omega_{\gamma P'} dt dE_\trg dV_{P'} dV_P} \br) \br] \ ,
\eeq
where we have integrated over the origin $P$ of the triggerons (expressed more naturally in terms of $s$ in our coordinates, \ie, $\dd V_P = \dd^3 q = \dd^3 s$), as well as their energy. The los-and-ROI integration is equivalent to integrating over the volume elements sourcing the signal but with $\mathcal{O}$ as the center of coordinates, \ie~$\dd V_{P'} = \dd^3 r = \dd \Omega_{\rm ROI} \, \dd\ell \, \ell^2$. The factor of $\ell^2$ here is cancelled by the flux dilution caused by distance $\ell$ between $P'$ and $\mathcal{O}$: the number of photons at $\mathcal{O}$ per unit energy per unit area, per interaction event, is $\ell^{-2} \, d^2 N_\gamma / dE_\gamma d\Omega_{\gamma P'}$. The $1/4 \pi$ overall coefficient in \Eq{eq:dPhidE} is standard in the indirect detection literature; it signifies the fraction of the celestial sphere at $P'$ crossed by the photons collected at $\mathcal{O}$ in the case of isotropic emission. However, this is already accounted for in the $\Omega_{\gamma P'}$ dependence of $d^2 N_\gamma / dE_\gamma d\Omega_{\gamma P'}$, as can be more easily seen in the general case of anisotropic emission. Furthermore, since this quantity is already properly normalized, we must correct the standard $1/4\pi$ prefactor with an extra factor of $4\pi$, shown inside the parentheses. For example, if each point $P'$ emits $\overline{N}$ photons isotropically per annihilation event, then $d N_\gamma / d\Omega_{\gamma P'} = \int\dd E_\gamma d^2 N_\gamma / dE_\gamma d\Omega_{\gamma P'}  = \overline{N} / 4\pi$. In the anisotropic case, the photon distribution $d^2 N_\gamma / dE_\gamma d\Omega_{\gamma P'}$ in \Eq{eq:dPhidE} must be evaluated at the coordinates $\Omega_{\gamma P'}$ that correspond to the direction of the observer $\mathcal{O}$ as seen from $P'$.

Without loss of generality, let us rotate our coordinates such that $P'$ lies along the vertical $\hat{\mathbf{k}}$ axis, as shown in \Fig{fig:geometry}. Our goal is to compute the two factors in \Eq{eq:d6N}, perform the integrals inside the square brackets of \Eq{eq:dPhidE}, and only afterwards do the los-and-ROI integrals over the observed locations $P'$. This can be done either by ``releasing'' $P'$ from the vertical axis and integrating as we let it move around the ROI {\it or}, alternatively, by integrating as we vary both $\abs{\mathbf{r}}$ and the apparent position of the observer in the celestial sphere $S^2_{P'}$, while keeping $P'$ along $\hat{\mathbf{k}}$.

\subsection{The annihilation rate}\label{app:B1}

Let us focus on the first factor of \Eq{eq:d6N}, the triggeron--pDM annihilation rate density $d^4 N_\gamma / dt dE_\trg dV_{P'} dV_P$. In general, this rate depends on the triggeron energy through the PBH emission rate spectrum and the annihilation cross section $\sigma_\ann(E_\trg)$.\footnote{For simplicity, we assume that the pDM is at rest, neglecting the subleading effects of the pDM halo velocity.} Axial symmetry, represented by the dotted blue circle in \Fig{fig:geometry}, guarantees that this rate depends only on $r \equiv \abs{\mathbf{r}}$, $s \equiv \abs{\mathbf{s}}$, and the relative angle between $\mathbf{r}$ and $\mathbf{s}$, which we parametrize by $u \equiv \mathbf{r} \cdot \mathbf{s} / rs$.
Generalizing the results of Ref. \cite{Agashe:2020luo} for our PBH-induced annihilations, the rate is
\bea
    \frac{d^4 N_\ann (r, s, u, E_\trg)}{dt dE_\trg dV_{P'} dV_P} & = & \frac{n_\PBH(q)}{4 \pi s^2} \, \frac{d^2 N_\trg^{1\PBH} (E_\trg)}{dt \, dE_\trg} \, \frac{d\mathcal{P}_\ann (r, s, u, E_\trg)}{d s} \ , \label{eq:ann_rate} \\
    \text{where} \quad \frac{d\mathcal{P}_\ann (r, s, u, E_\trg)}{d s} & = & n_\pDM(r) \, \sigma_\ann(E_\trg) \, \exp \bl[ - s \, \sigma_\ann(E_\trg) \, \int\limits_0^1 \dd z \, n_\pDM(\abs{\mathbf{r} + z \mathbf{s}}) \br] \label{eq:pann}
\eea
is the triggeron annihilation probability per unit distance. Here $d^2N_\trg^{1\PBH}(E_\trg)/dt dE_\trg$ is the spectrum of the triggerons emitted by one PBH, and is given by \Eq{eq:hawk_rad}. The PBH number density at $P$, a distance $q \equiv \abs{\mathbf{q}}$ from $O$, is $n_\PBH(q)$, while the pDM number density at the annihilation point $P'$ is $n_\pDM(r)$. Note that we have omitted the arguments of $q$ showing its $\mathbf{r}$- and $\mathbf{s}$-dependence: $q = q(\mathbf{r}, \mathbf{s}) = \sqrt{r^2 + s^2 + 2 r s u}$. The exponential in \Eq{eq:pann} accounts for the probability that the triggeron has survived all the way from its birth-by-Hawking at $P$ until its death-by-annihilation at $P'$ \cite{Agashe:2020luo}; the vector $\mathbf{r} + z \mathbf{s}$, with $0 \leq z \leq 1$, parametrizes the triggeron's location along its path of flight on the $P'P$ line segment. Clearly, $\abs{\mathbf{r} + z \mathbf{s}} = q(\mathbf{r}, z \mathbf{s}) = \sqrt{r^2 + s^2 z^2 + 2 r s u z}$. Finally, the factor of $1/4 \pi s^2$ in \Eq{eq:ann_rate} is due to the dilution at $P'$ of the triggeron flux emitted at $P$.

\subsection{The photon distribution}\label{app:B2}

We now turn our attention to the second factor in \Eq{eq:d6N}, the angular and energy distribution of the photons created at $P'$, namely $d^2 N_\gamma(E_\gamma, E_\trg, \Omega_{\gamma P'}) / dE_\gamma d\Omega_{\gamma P'}$. As mentioned above, this distribution is generally {\it not} spherically symmetric, hence its dependence on the coordinates $\Omega_{\gamma P'}$ of $S^2_{P'}$. This angular dependence arises both from the fact that in general $P' \neq O$, and from the angular distribution of the annihilation and decay products. This distribution is in general very complicated, but we can find analytic expressions for it in special cases and specific reference frames. We consider three distinct frames: the mediator's rest frame (RF), the center-of-mass frame (CF) of each triggeron--pDM annihilation event, and the galactic frame (GF) which is the same as that of the observer, since we neglect the relatively small orbital motions of the Sun and the Earth.

Let us begin by describing the kinematics of the annihilation and decay processes. As in our BaDM and DTDM models, and as spelled out in \Eq{eq:steps}, we focus on 2-to-2 annihilations and two-body decays:
\bea\label{eq:app_steps}
    \text{annihilations: } &&\!\!\!\!\!\!\! \text{ triggeron } + \text{ pDM } \to \text{ mediator } + X \ ; \nonumber\\
    \text{decays: } && \!\!\!\!\!\!\! \text{ mediator } \to \gamma + Y \ ,
\eea
where $X$ and $Y$ stand for additional particles in these processes, with masses $m_X$ and $m_Y$, respectively. For example, in the BaDM model $X = \phi$, and $Y = \gamma$, whereas in the DTDM model $X = W'$ and $Y = \chi_1$; see \Eqs{eq:steps_badm}{eq:steps_dtdm}. Energy-momentum conservation ensures that the energy spectrum of the mediator and of the photon in \Eq{eq:app_steps} are monochromatic. In order to find their angular distribution, let us assume for simplicity that the spins of the triggerons and of the pDM are randomly distributed at every point in the halo. This should be true for sufficiently large volumes $dV$ and for sufficiently long time intervals $dt$. Consequently, there is no {\it net average} spin in the annihilation and decay events involved, and thus no preferred direction in the outgoing mediator and photon distributions at $P'$. The only preferred direction is that of the total momentum vector in the GF, which is the same as that of the triggeron's momentum, $-\hat{\mathbf{s}}$, and it only plays a role in the GF. Therefore, the monochromatic distributions of the photon in the RF and of the mediator in the CF are also isotropic.

In general, a monochromatic and isotropic distribution of a particle of mass $m$ and energy $E^{\rm [F1]}$ in a reference frame F1 can be written in terms of a Dirac delta function:
\beq\label{eq:dNF1}
    \bl. \frac{d^2 N}{dE d\Omega} \br\vert^{\rm [F1]} = \frac{\overline{c}}{4 \pi} \delta \bl( E^{\rm [F1]} - E_*^{\rm [F1]} \br) \ ,
\eeq
where $E_*^{\rm [F1]}$ is the monochromatic energy of the outgoing particle and is determined purely by kinematic considerations, and the constant $\overline{c}$ denotes the multiplicity of the particles in question. It is evident that this distribution is properly normalized: $\int \dd E \dd \Omega \bl. d^2 N / d E d \Omega \br\vert^{\rm [F1]} = \overline{c}$. For example, in the BaDM model there are $\overline{c}_\med = \overline{c}_\phi = 2$ $\phi$ mediators per annihilation event and $\overline{c}_\gamma = 2$ photons per decay event, whereas in the DTDM model there are $\overline{c}_\med = \overline{c}_2 = 1$ $\chi_2$ mediators per annihilation and $\overline{c}_\gamma  =1$ photons per decay.

Now let us suppose that F1 is moving with a velocity $\mathbf{v}_{\rm F1}^{\rm [F2]} \equiv v \, \hat{\mathbf{v}}$ with respect to another frame F2. Our task is to find the particle distribution in the new frame F2, in terms of the F2 kinematic variables:
\beq
    \bl. \frac{d^2 N}{dE d\Omega} \br\vert^{\rm [F2]} = \abs{\mathbf{J}} \, \frac{\overline{c}}{4 \pi} \delta \bl( E^{\rm [F1]}(E^{\rm [F2]}, \Omega^{\rm [F2]}) - E_*^{\rm [F1]} \br) \ ,
\eeq
where $\abs{\mathbf{J}}$ is the Jacobian of the change of variables between frames $(E^{\rm [F1]} , \Omega^{\rm [F1]}) \to (E^{\rm [F2]} , \Omega^{\rm [F2]})$, and $E^{\rm [F1]}(E^{\rm [F2]}, \Omega^{\rm [F2]})$ is $E^{\rm [F1]}$ written in terms of the F2 variables. To find the 4-momentum in F1 in terms of the 4-momentum in F2 we need to perform an active Lorentz transformation with boost $\boldsymbol{\beta} \equiv - \beta \,\hat{\mathbf{v}}$ (and $\beta = v$), and a corresponding Lorentz factor $\gamma = 1/\sqrt{1-\beta^2}$. This yields
\bea
    E^{\rm [F1]} & = & \gamma E^{\rm [F2]} - \gamma \beta \, p^{\rm [F2]} \cos\theta^{\rm [F2]} \label{eq:EF1} \\
    \text{and} \quad p^{\rm [F1]} \cos\theta^{\rm [F1]} & \equiv & \gamma p^{\rm [F2]} \cos\theta^{\rm [F2]} - \gamma \beta \, E^{\rm [F2]} \ ,\label{eq:pF1}\\
    \text{where} \quad p & \equiv & \sqrt{E^2 - m^2} \\
    \text{and} \quad \cos\theta & \equiv & \hat{\mathbf{p}} \cdot \hat{\mathbf{v}} \ .
\eea
Plugging \Eq{eq:EF1} in \Eq{eq:pF1} allows us to find $\cos\theta^{\rm [F1]}$ as a function of $E^{\rm [F2]}$ and $\cos\theta^{\rm [F2]}$. Since $\dd \Omega = \dd \phi \dd(\cos\theta)$ (defining the polar angle $\theta$ with respect to $\hat{\mathbf{v}}$) and $\dd \phi^{\rm [F1]} = \dd \phi^{\rm [F2]}$, we find
\beq
    \abs{\mathbf{J}} = \frac{p^{\rm [F2]}}{\sqrt{\gamma^2 \bl( E^{\rm [F2]} - \beta \, p^{\rm [F2]} \cos\theta^{\rm [F2]} \br)^2 - m^2}}
\eeq
and, upon some manipulation,
\beq\label{eq:dNF2}
    \bl. \frac{d^2 N}{dE d\Omega} \br\vert^{\rm [F2]} = \frac{\overline{c}}{4 \pi \, \gamma \beta p_*^{\rm [F1]}} \delta\bl( \cos\theta^{\rm [F2]} - \frac{\gamma E^{\rm [F2]} - E_*^{\rm [F1]}}{\gamma \beta p^{\rm [F2]}} \br) \ ,
\eeq
where we have evaluated the Jacobian on the value of $\cos\theta^{\rm [F2]}$ picked by the Dirac delta. It is evident that in F2 the particle distribution is no longer monochromatic nor isotropic, but that the energy of the particles is in one-to-one correspondence to the emission angle $\theta^{\rm [F2]}$. Indeed, upon integration over the angular variables we get a box-shaped spectrum:
\bea
    \bl. \frac{d N}{d E} \br\vert^{\rm [F2]} & = & \frac{\overline{c}}{\Delta E^{\rm [F2]}} \mathrm{Box}\bl[E^{\rm [F2]} ; \, \overline{E}^{\rm [F2]}, \, \Delta E^{\rm [F2]} \br] \ , \label{eq:box}\\
    \text{where} \quad \overline{E}^{\rm [F2]} & \equiv & \gamma E_*^{\rm [F1]} \\
    \text{and} \quad \Delta E^{\rm [F2]} & \equiv & 2 \gamma \beta p_*^{\rm [F1]} \ ,
\eea
and $\mathrm{Box}(x; \, \overline{x}, \, \Delta x)$ is the unit box function centered at $\overline{x}$ and with a width $\Delta x$. Clearly, the largest (smallest) particle energies in F2 correspond to those partcles in F1 that had momentum parallel (antiparallel) to $\mathbf{v}$, \ie, $\theta^{\rm [F2]} = 0$ ($\theta^{\rm [F2]} = \pi$). As a sanity check, we can see that further integrating over the energy yields $\int \dd E^{\rm [F2]} \, \bl. dN/dE \br\vert^{\rm [F2]} = \overline{c}$, as expected.

Let us bring this back to the annihilation and decay processes in \Eq{eq:app_steps}. The absence of a preferred direction in the RF means that the RF photon distribution {\it per decay event} is monochromatic and isotropic:
\bea
    \bl. \frac{d^2 N_\gamma^{1\dec}}{d E_\gamma d\Omega_\gamma} \br\vert^{\rm [RF]} & = & \frac{\overline{c}_\gamma}{4\pi} \delta\bl( E_\gamma^{\rm [RF]} - E_{\gamma*}^{\rm [RF]} \br) \ , \label{eq:dNg_RF}\\
    \text{with} \quad E_{\gamma*}^{\rm [RF]} & \equiv & \frac{m_\med^2 - m_Y^2}{2 m_\med} \ . \label{eq:Egstar}
\eea
For example, in the BaDM model $E_{\gamma*}^{\rm [RF]} = m_\phi / 2$; whereas in the DTDM model $E_{\gamma*}^{\rm [RF]} = \Delta m / (1 + \Delta m / (2m)) \simeq \Delta m$, with $\Delta m \equiv m_2 - m_1$ and $m \equiv (m_1 + m_2) / 2 \gg \Delta m$. By the same reasoning, the absence of a preferred direction in the CF means that the CF mediator distribution {\it per annihilation event} is also monochromatic and isotropic:
\bea
    \bl. \frac{d^2 N_\med}{d E_\med d\Omega_\med} \br\vert^{\rm [CF]} & = & \frac{\overline{c}_\med}{4\pi} \delta\bl( E_\med^{\rm [CF]} - E_{\med*}^{\rm [CF]} \br) \ , \label{eq:dNmed_CF}\\
    \text{where} \quad E_{\med*}^{\rm [CF]} & \equiv & \frac{s_M + m_\med^2 - m_X^2}{2 \sqrt{s_M}} \ , \label{eq:Emstar} \\
    \text{and} \quad s_M & \equiv & 2 E_\trg m_\pDM + m_\pDM^2 + m_\trg^2 \equiv \bl( E_{\rm tot}^{\rm [CF]} \br)^2 \label{eq:sM}
\eea
is the Mandelstam $s$-variable (equal to the square of the total energy in the CF, $E_{\rm tot}^{\rm [CF]}$) and is not to be confused with the distance $s$ between $P$ and $P'$.

In order to obtain the photon distribution {\it per annihilation event} in the GF $d^2 N_\gamma / dE_\gamma d\Omega_{\gamma P'}$, needed in \Eq{eq:d6N}, we need to first determine the velocity of the RF in the GF {\it for each decay event}, write the equivalent of \Eq{eq:dNF2} for \Eq{eq:dNg_RF} with F1=RF and F2=GF, and then integrate over the distribution of decay events. The velocity in question is of course that of the mediator in the GF, \ie, $\mathbf{v}_{\rm RF} = \mathbf{v}_\med$; the required boost, which we denote by $\boldsymbol{\beta}_\dec = - \mathbf{v}_\med$, is given by the mediator's kinematics:
\beq\label{eq:dec_boost}
    \gamma_\dec \equiv E_\med / m_\med \ , \qquad \beta_\dec = \sqrt{1 - \gamma_\dec^{-2}} \ .
\eeq
The magnitude {\it and} direction of $\boldsymbol{\beta}_\dec$ is distributed according to the energy and angular distribution of the outgoing mediators in each annihilation event, which in turn follows the equivalent of \Eq{eq:dNF2} for \Eq{eq:dNmed_CF}. In this case, the boost in question, which we denote by $\boldsymbol{\beta}_\ann$, is given by the velocity of the CF in the GF:
\beq\label{eq:ann_boost}
    \gamma_\ann \equiv \frac{E_{\rm tot}}{E_{\rm tot}^{\rm [CF]}} = \frac{m_\pDM + E_\trg}{\sqrt{s_M}} \ , \qquad \boldsymbol{\beta}_\ann = - \mathbf{v}_{\rm CF} = \sqrt{1 - \gamma_\ann^{-2}} \, \hat{\mathbf{s}} \ .
\eeq
since the CF moves in the same direction as the triggeron with respect to the GF, namely $- \hat{\mathbf{s}}$.

The final result for the angular and energy distribution of the photons per annihilation event at $P'$ in the GF is therefore:
\bea
    \frac{d^2 N_\gamma}{dE_\gamma d\Omega_{\gamma P'}} & = & \int \dd E_\med \int \dd \Omega_\med \ \frac{d^2 N_\med}{dE_\med d \Omega_\med} \, \abs{\mathbf{J}'} \frac{d^2 N_\gamma^{1\dec}}{dE_\gamma d \Omega_\gamma} \nonumber\\
    & = & \frac{\overline{c}_\gamma \overline{c}_\med}{2 \pi \, \Delta E_\gamma} \int \frac{\dd E_\med}{\Delta E_\med} \int \frac{\dd \Omega_\med}{2 \pi} \ \delta\bl( \hat{\mathbf{p}}_\med \cdot \hat{\mathbf{v}}_{\rm CF} - \frac{E_\med - E_{\med*}^{\rm [CF]}}{\Delta E_\med / 2} \frac{p_{\med*}^{\rm [CF]}}{p_\med} \br)
    \nonumber\\
    && \times \ \abs{\mathbf{J}'} \, \delta \bl( \hat{\mathbf{p}}_\gamma \cdot \hat{\mathbf{p}}_\med - \frac{E_\gamma - E_{\gamma *}^{\rm [RF]}}{\Delta E_\gamma / 2} \frac{E_{\gamma*}^{\rm [RF]}}{E_\gamma} \br)
    \label{eq:photon_dist}\\
    \text{where} \quad \overline{E}_\med & = & \gamma_\ann E_{\med*}^{\rm [CF]} \ , \quad \Delta E_\med \equiv 2 \gamma_\ann \beta_\ann p_{\med*}^{\rm [CF]} \label{eq:mediator_en} \ ; \\
    \overline{E}_\gamma & = & \gamma_\dec E_{\gamma*}^{\rm [RF]}  \ , \quad \Delta E_\gamma \equiv 2 \gamma_\dec \beta_\dec E_{\gamma*}^{\rm [RF]} \label{eq:photon_en} \ .
\eea
Note that $\Omega_\gamma \neq \Omega_{\gamma P'}$; $\abs{\mathbf{J}'}$ is the Jacobian from the $\Omega_\gamma \to \Omega_{\gamma P'}$ change of coordinates. This change is necessary because the polar angle $\theta_\gamma$ in $\Omega_\gamma$ is defined with respect to the direction of motion of the RF with respect to the GF, namely the variable $\hat{\mathbf{p}}_\med$; indeed, see how $\Omega$ in the general formula of \Eq{eq:dNF2} is defined with respect to $\hat{\mathbf{v}}$, the motion of F1 in F2. In other words, the definition of $\Omega_\gamma$ {\it changes} when integrating over the direction of motion of the mediator, $\dd\Omega_\med$. A fixed definition of the photon angular coordinates is necessary, and thus the reason for the Jacobian $\abs{\mathbf{J}'}$, which in general depends on $\Omega_\med$. Note, however, that this subtlety is not an issue for the $\theta_\med$ polar angle of the mediator itself, which is defined with respect to the fixed $-\hat{\mathbf{s}}$ direction: $\hat{\mathbf{p}}_\med \cdot \hat{\mathbf{v}}_{\rm CF} = - \hat{\mathbf{p}}_\med \cdot \hat{\mathbf{s}} \equiv \cos\theta_\med$.

\subsection{Simplifying Limits}\label{app:B3}

We have arrived to the most general expressions for the factors in \Eq{eq:d6N}, given by \Eqs{eq:ann_rate}{eq:photon_dist}, and which are necessary to compute the differential flux, \Eq{eq:dPhidE}. While \Eq{eq:ann_rate} seems complicated enough, \Eq{eq:photon_dist} is positively ghastly, its angular dependence being far too intricate. Thankfully, we can take two limit cases where our scenario is simplified to such an extent that semi-analytic results can be obtained with only a modest effort.

\subsubsection{The local limit}

The first case is the {\it local limit}, when the annihilation probability per unit distance in \Eq{eq:pann} is so large that the triggerons are annihilated in the vicinity of their birthplace at $P$. More concretely, let us write the halo number density of the $i$-th DM species (either pDM or PBH) in terms of a universal dimensionless profile density $\eta(\hat{r})$ as follows:
\beq
    n_i(r) = \frac{\rho_i(r)}{m_i} = \frac{f_i \rho_0}{m_i} \eta(\hat{r}) \ , \quad \text{with} \quad \hat{r} \equiv r/r_s \ ,
\eeq
and where $m_i$ is the mass of the $i$-th species, $f_i$ its halo mass fraction, and $\rho_0$ and $r_s$ are the DM density and scale radius parameters describing the total DM density profile $\rho(r)$. For example the Navarro-Frenk-White (NFW) profile \cite{Navarro:1994hi,Navarro:1995iw,Navarro:1996gj}, is $\eta(\hat{r}) = 1/\hat{r}(1+\hat{r})^2$, with $\rho_0 = 0.838~\GeV \cm^{-3}$, and $r_s = 11~\kpc$ \cite{deSalas:2019pee}. Note that we use hatted quantities $\hat{x} = x/r_s$ to denote length scales rescaled by $r_s$; these are not to be confused with our notation for unit vectors, which are typeset in boldface, \eg~$\hat{\mathbf{v}}$. Furthermore, as in the main text, we define the {\it annihilation coefficient} $\Lambda$ as the inverse of the typical triggeron mean-free-path in units of $r_s$, \ie~$\Lambda(E_\trg) \equiv f_\pDM \rho_0 \sigma_\ann(E_\trg) r_s / m_\pDM$. Then the local limit corresponds to $\Lambda \gg 1$ or, formally,
\beq\label{eq:local_limit}
    \lim\limits_{\Lambda \eta \to \infty} \, \frac{d \mathcal{P}_\ann}{d s} = \delta(s) \ ;
\eeq
and therefore $P' = P$ \cite{Agashe:2020luo}. For succinctness, we have omitted the dependence of $\Lambda$ on $E_\trg$.

In the local limit, the preferred direction $-\hat{\mathbf{s}}$, induced by the triggeron flux from $P$ to $P'$, is gone, and spherical symmetry is restored by default. Said differently, the photons emitted at $P'$ come from mediator particles boosted in {\it all} directions equally. This means that the photon distribution $d^2 N_\gamma / dE_\gamma d\Omega_{\gamma P'}$ is isotropic and can be written as
\bea
    \frac{d^2 N_\gamma}{dE_\gamma d\Omega_{\gamma P'}} = \frac{1}{4 \pi} \int \dd\Omega_{\gamma P'} \ \frac{d^2 N_\gamma}{dE_\gamma d\Omega_{\gamma P'}} & = & \frac{1}{4 \pi} \int \dd \Omega_{\gamma P'} \int \dd E_\med \int \dd \Omega_\med \ \frac{d^2 N_\med}{dE_\med d \Omega_\med} \, \abs{\mathbf{J}'} \frac{d^2 N_\gamma^{1\dec}}{dE_\gamma d \Omega_\gamma} \nonumber\\
    & = & \frac{1}{4 \pi} \int \dd \Omega_\gamma \int \dd E_\med \int \dd \Omega_\med \ \frac{d^2 N_\med}{dE_\med d \Omega_\med} \, \frac{d^2 N_\gamma^{1\dec}}{dE_\gamma d \Omega_\gamma} \ ,
\eea
where in the last equality the integration over $\dd\Omega_{\gamma P'}$ allows us to get rid of the Jacobian $\abs{\mathbf{J}'}$ by reverting to the $\Omega_\gamma$ coordinates. This is exactly \Eq{eq:photon_dist} integrated over all the possible photon angles. Finally, this means that
\bea
    \frac{d^2 N_\gamma}{dE_\gamma d\Omega_{\gamma P'}} & = & \frac{1}{4 \pi} \frac{d N_\gamma}{d E_\gamma} \ , \label{eq:isotropic_spectrum}\\
    \text{where} \quad \frac{d N_\gamma}{d E_\gamma} & = &
    \frac{\overline{c}_\gamma \overline{c}_\med}{\Delta E_\gamma} \int \frac{\dd E_\med}{\Delta E_\med} \ \mathrm{Box} \bl[ E_\med; \overline{E}_\med, \Delta E_\med \br] \, \mathrm{Box} \bl[ E_\gamma; \overline{E}_\gamma, \Delta E_\gamma \br] \ , \label{eq:photon_spectrum}
\eea
and $\overline{E}_{\med/\gamma}$ and $\Delta E_{\med/\gamma}$ are defined as in \Eqs{eq:mediator_en}{eq:photon_en}, with the help of Eqs.~\ref{eq:Egstar}, \ref{eq:Emstar}-\ref{eq:ann_boost}. The integral in \Eq{eq:photon_spectrum} can be done numerically; as expected from the convolution of two box functions, and as promised in \Sec{sec:processes}, it yields a trapezoidal spectrum.

Using \Eq{eq:local_limit} in \Eqs{eq:ann_rate}{eq:dPhidE}, we arrive at \Eq{eq:dPhidE_simplest}, namely the differential flux in the local limit:
\bea
    \frac{d \Phi_\gamma}{d E_\gamma} & = & \frac{1}{4 \pi} \frac{f_\PBH}{M_\PBH}  \, J_\dec \, \int \dd E_\trg \ \frac{d^2 N^{\rm 1 PBH}_\trg (E_\trg)}{dt dE_\trg} \, \frac{d N_\gamma (E_\trg, E_\gamma)}{d E_\gamma} \ , \label{eq:dPhi_local} \\
    \text{where} \quad J_\dec & \equiv & \int\limits_{\rm ROI} \dd\Omega \int\limits_{\rm los} \dd \ell \ \rho(r) \label{eq:Jdec_local}
\eea
is the $J$-factor for decaying DM common in the indirect detection literature, and $dN_\gamma / dE_\gamma$ is given by \Eq{eq:photon_spectrum}. Throughout our paper we are interested in a ROI spanning an angle $\theta_{\rm ROI}$ around the GC, which has a NFW $J$-factor given by
\bea
    J_\dec(\theta_{\rm ROI}) & = & \rho_0 r_s \ 2\pi \int\limits_{0}^{\theta_{\rm ROI}}\dd\theta \sin\theta \, \int\limits_{0}^{\hat{\ell}_{\max} (\theta)} \dd\hat{\ell} \ \eta(\hat{r}(\hat{\ell}, \theta)) \ , \label{eq:Jfac_cone} \\
    \text{where} \quad r(\ell, \theta) & = & \sqrt{\ell^2 + d_\odot^2 - 2 \ell d_\odot \cos\theta} \ , \label{eq:r_elltheta} \\
    \text{and} \quad \ell_{\max}(\theta) & \equiv & d_\odot \cos\theta + \sqrt{r_{200}^2 - d_\odot^2 \sin^2\theta} \ , \label{eq:ell_max}
\eea
and we have once again used the hatted notation in \Eq{eq:Jfac_cone}. Here $d_\odot$ is the distance from the Sun to the GC, and $r_{200}$ is the commonly used definition for the virial radius of the DM halo, \ie, the radius at which the average DM density $\overline{\rho} \equiv 4 \pi \int\limits_{0}^{r_{200}} \dd r r^2 \ \rho(r) /(4 \pi r_{200}^3 / 3)$ is 200 times the background critical density of the Universe, $\rho_{\rm crit} \equiv 3 H_0^2 / (8 \pi G_N)$, where $H_0$ is the Hubble parameter. For $H_0 = 67.66~\km \sec^{-1} \Mpc^{-1}$ \cite{Planck:2018vyg}, $d_\odot = 8.122~\kpc$ \cite{GRAVITY:2018ofz}, and our NFW parameters, we get $r_{200} \approx 189~\kpc$, and $J_\dec(5\degree) \approx 3.8 \times 10^{24}~\MeV \cm^{-2}$ for a ROI of $\theta_{\rm ROI} = 5 \degree$ ($\Omega_{\rm ROI} \approx 0.024~\mathrm{sr}$).

\subsubsection{The distant-sphere limit}

Let us now consider finite $\Lambda$, in which case $P \neq P'$ in general and the triggeron--pDM annihilations are necessarily non-local. There is an axial symmetry in our scenario that somewhat simplifies things. Upon integration over the azimuthal angle $\varphi_{\hat{\mathbf{s}}}$ in the differential volume $\dd V_P = \dd^3 s$ of \Eq{eq:dPhidE}, represented by the blue dashed circle in \Fig{fig:geometry}, the photon distribution $\int \dd\varphi_{\hat{\mathbf{s}}} \ d^2 N_\gamma / dE_\gamma d\Omega_{\gamma P'}$ will in the end depend only on the angle between $\hat{\mathbf{s}}$ and $\hat{\mathbf{r}} = \hat{\mathbf{k}}$ (which is the same for all points $P$ in the blue circle). However, this is not enough to obtain a simple expression for \Eq{eq:photon_dist}.

Nevertheless, simplicity is restored in the second special case we consider, which we call the {\it distant-sphere limit}. In this limit, the ROI emitting the photons we wish to observe is well described by a spherical ball whose radius is much smaller than its distance to the observer. More concretely, if we denote the volume in question as the ball $\mathbb{B}_R$ of radius $R$, all the points within it are approximately at the same distance $\ell \approx d \gg R$ from the observer. This is the case of signals coming from the dwarf spheroidal galaxies orbiting the Milky Way as well as other distant galaxies. It is also the case of a sufficiently small core ROI in the center of the Milky Way, distinct from the more commonly considered ROI spanning a few degrees from the GC, which are conical in shape.

The advantage of the distant-sphere limit is that spherical symmetry is restored. Indeed, the angular part of the $\dd V_{P'}$ integral in \Eq{eq:dPhidE} for fixed $r$ over the spherical ROI amounts to moving $P'$ around the red dashed sphere in \Fig{fig:geometry}. All the points in this sphere contribute with exactly the same photon distribution $d^2 N_\gamma / d E_\gamma d \Omega_{\gamma P'}$ and are at approximately the same distance $d$ from the observer. Crucially, however, the observer sees different solid angles $\Omega_{\gamma P'}$ of this photon distribution for each point in the red dashed sphere. In other words, integration over the angular part of $\dd V_{P'}$ at fixed $r$ is {\it the same} as integrating over all $\Omega_{\gamma P'}$ in the photon distribution. As a result the photon spectrum can be taken to be {\it effectively} isotropic and we can use \Eqs{eq:isotropic_spectrum}{eq:photon_spectrum}, just as in the local limit.

\begin{figure}
\centering
\includegraphics[width=0.6\linewidth]{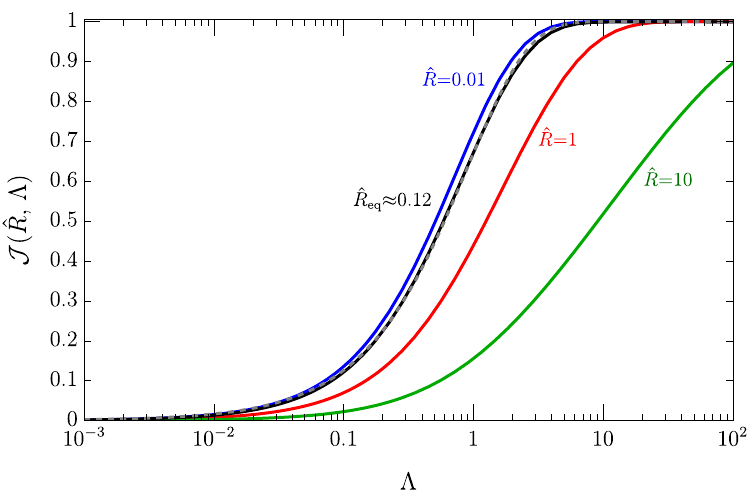}
\caption{Suppression factor of the photon signal in the distant-sphere limit, $\mathcal{J}(\hat{R}, \Lambda)$, given by \Eq{eq:curlyJ}. We plot its dependence as a function of $\Lambda$ for a handful of values of $\hat{R}$. In black, we plot $\mathcal{J}_{\rm eq}(\Lambda) \equiv \mathcal{J}(\hat{R}_{\rm eq}, \Lambda)$ for $\hat{R}_{\rm eq} \approx 0.116$, which is the suppression factor for the equivalent ball $\mathbb{B}_{R_{\rm eq}}$ of radius $R_{\rm eq} = \hat{R}_{\rm eq} \, r_s$ that gives the same $J$-factor as the conical ROI $5\degree$ around the GC. The dashed gray line is the fit $1 - \exp(-a \Lambda^b)$, with $a = 1.10$, and $b = 0.915$, to $\mathcal{J}_{\rm eq}(\Lambda)$.}
\label{fig:curlyJ}
\end{figure}

Resorting once again to the hatted notation for dimensionless length scales, and after some straightforward manipulations, we find that the differential photon flux in the distant-sphere limit is given by
\bea
    \frac{d \Phi_\gamma}{d E_\gamma} & = & \frac{1}{4 \pi} \frac{f_\PBH}{M_\PBH} J_\dec (\mathbb{B}_R) \, \int \dd E_\trg \ \frac{d^2 N_\trg^{1\PBH}}{dt \, d E_\trg} \frac{d N_\gamma}{d E_\gamma} \ \mathcal{J}(\hat{R}, \Lambda) \ , \label{eq:dPhi_sphere} \\
    \text{where} \quad \mathcal{J}(\hat{R}, \Lambda) & \equiv & \frac{1}{\mu(\hat{R})} \int\limits_0^{\hat{R}} \dd \hat{r} \, \hat{r}^2 \int\limits_0^{\hat{R}} \dd \hat{s} \, \frac{1}{2} \int\limits_{-1}^{1} \dd u \, \eta(\hat{q}(\hat{r}, \hat{s}, u)) \ \Lambda \, \eta(\hat{r}) \exp\bl[ - \Lambda \, \hat{s} \ \int\limits_0^1 \dd z \ \eta(\hat{q}(\hat{r}, z \hat{s}, u)) \br] \ , \label{eq:curlyJ}\\
    \mu(\hat{R}) & \equiv & \int\limits_0^{\hat{R}} \dd \hat{r} \, \hat{r}^2 \eta(\hat{r}) = \ln\bl( 1 + \hat{R} \br) - \frac{\hat{R}}{1+\hat{R}} \ , \label{eq:muFn} \\
    \text{and} \quad J_\dec (\mathbb{B}_R) & \equiv & \frac{4 \pi r_s^3 \rho_0}{d^2} \mu(\hat{R}) \ ,\label{eq:Jdec_ball}
\eea
having assumed an NFW profile for $\eta(\hat{r})$. Here $J_\dec (\mathbb{B}_R)$ is the DM decay $J$-factor for the $\mathbb{B}_R$ ball ROI, $\hat{q}(\hat{r}, \hat{x}, u) \equiv \sqrt{\hat{r}^2 + \hat{x}^2 + 2 \hat{r} \hat{x} u}$, and $dN_\gamma / dE_\gamma$ is once again given by \Eq{eq:photon_spectrum}. Note that in general $\Lambda$ depends on $E_\trg$ and thus $\mathcal{J}(\hat{R}, \Lambda)$ cannot be pulled outside the integral in \Eq{eq:dPhi_sphere}. Equation~(\ref{eq:curlyJ}) can only be computed numerically; however it can be shown that in the local limit $\mathcal{J}(\hat{R}, \Lambda \to \infty) \to 1$, and that $\mathcal{J}(\hat{R}, \Lambda \to 0) \to 0$, as we could have expected. This behavior is clearly seen in \Fig{fig:curlyJ}, where we show $\mathcal{J}(\hat{R}, \Lambda)$ as a function of $\Lambda$ for several values of $\hat{R}$. This means that $\mathcal{J}(\Lambda)$ can be interpreted as an effective suppression of the $J$-factor, and consequently the overall signal, due to the triggeron--pDM annihilation probability not being exactly one. In other words, since the triggerons have a finite mean-free-path, a fraction $\sim 1 - \mathcal{J}$ of them escapes the galaxy without annihilating.

\subsection{Summary: the impact of finite $\Lambda$}\label{app:B4}

The differential photon flux $d\Phi_\gamma / dE_\gamma$ is equal to \Eq{eq:dPhi_local} in the local limit $\Lambda \to \infty$, while for finite $\Lambda$ it is given by \Eq{eq:dPhi_sphere} in the distant-sphere limit. Therein lies the difficulty in calculating the amplitude of the photon signal coming from the Milky Way: Unless $\Lambda$ is very large, we cannot use \Eq{eq:dPhi_local} and yet, since we as observers are within the galactic halo, the distant-sphere limit is not valid either. Indeed, a ROI around the GC, like the one considered in the main text of this paper, is not spherical but conical in shape.

\begin{figure}
\centering
\includegraphics[width=0.7\linewidth]{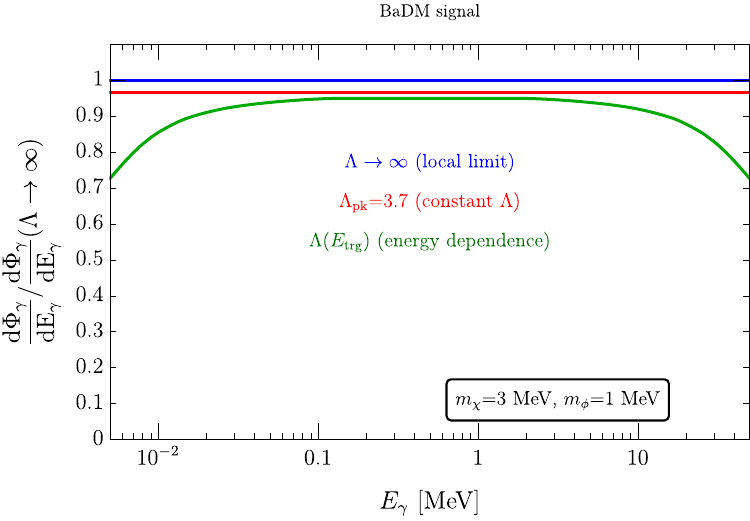}
\includegraphics[width=0.7\linewidth]{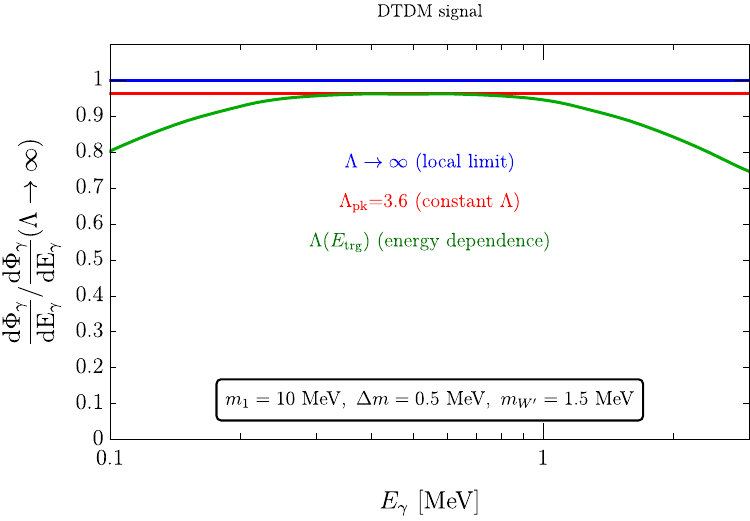}
\caption{The differential photon flux $d \Phi_\gamma / d E_\gamma$ of the DS signal in the distant-sphere limit, given by \Eq{eq:dPhi_sphere} for finite $\Lambda$, compared to the same flux in the local limit ($\Lambda \to \infty$, blue line). We have chosen $\hat{R} = \hat{R}_{\rm eq} \approx 0.116$, the radius of the equivalent ball that gives the same $J$-factor as the $5\degree$ ROI around the GC. We plot the BaDM (DTDM) signal in the top (bottom) panel, for our benchmark masses $m_\chi = 3~\MeV$ and $m_\phi = 1~\MeV$ ($m_1 = 10~\MeV$, $\Delta m = 0.5~\MeV$, and $m_{W'} = 1.5~\MeV$). For BaDM, we have chosen $M_\PBH = 10^{15}~\gr$ (\ie, $T_\PBH \approx 11~\MeV$), while for DTDM, we chose $M_\PBH = 10^{16}~\gr$ (\ie, $T_\PBH \approx 1.1~\MeV$). The red lines correspond to the case of constant $\Lambda = \Lambda_{\rm pk}$ ($\approx 3.7$ for BaDM and $\approx 3.6$ for DTDM; see \Eqs{eq:Lambda_badm}{eq:Lambda_dtdm}), defined at the peak triggeron energy of the Hawking radiation $E_\trg = E_{\rm pk}$ ($\approx 23~\MeV$ for the BaDM $\anti{\chi}$ triggerons, and $\approx 6~\MeV$ for the DTDM $W'$ triggerons). The green lines correspond to the energy-dependent case of $\Lambda(E_\trg)$, assuming $\sigma_\ann(E_\trg) \propto p_\med^{\rm [CF]} / \bl( s_M \, p_\pDM^{\rm [CF]} \br)$ in \Eq{eq:Lambda_pk}. The effect of the energy dependence is mostly limited to a suppression of the lower- and higher-energy ends of the signal, which come from the most boosted photons.}
\label{fig:Lambda}
\end{figure}

Nevertheless, we can estimate this amplitude in terms of the suppression factor $\mathcal{J}$ of \Eq{eq:curlyJ}. The simplest way to do this is to define an {\it equivalent ball} $\mathbb{B}_{R_{\rm eq}}$ whose $J$-factor in \Eq{eq:Jdec_ball} is the same as that of the ROI in question, and then compute $\mathcal{J}_{\rm eq}(\Lambda) \equiv \mathcal{J}(\hat{R}_{\rm eq}, \Lambda)$ for that radius and a given $\Lambda$. For example, for our NFW parameters, a ROI of $5\degree$ around the GC has an equivalent ball of radius $R_{\rm eq} \approx 1.28~\kpc$ ($\hat{R}_{\rm eq} \approx 0.116$). The $\mathcal{J}_{\rm eq}(\Lambda)$ suppression, shown as a thick black line in \Fig{fig:curlyJ}, can be fit to using the semi-analytic function $1 - \exp(-a \Lambda^b)$, where $a = 1.10$, and $b = 0.915$ (gray dashed line). For $\Lambda = 1$ we estimate a suppression factor of $\mathcal{J} \approx 0.67$ (\ie, a signal 33\% smaller than its value in the local limit), whereas for our $\Lambda \approx 4$ benchmark (see \Eqs{eq:Lambda_badm}{eq:Lambda_dtdm}) we find $\mathcal{J} \approx 0.96$ (\ie, only a 4\% smaller signal). In \Fig{fig:Lambda} we show the impact that finite $\Lambda$ has on the differential photon flux of the signal from the BaDM and DTDM models.

Finally, let us say a few words about the dependence of $\Lambda$ on the triggeron energy $E_\trg$, which arises from that of the triggeron--pDM annihilation cross section $\sigma(E_\trg)$. In general, we can parametrize $\Lambda(E_\trg)$ as
\beq\label{eq:Lambda_pk}
    \Lambda(E_\trg) \equiv \Lambda_{\rm pk} \frac{\sigma_\ann(E_\trg)}{\sigma_\ann(E_{\rm pk})} \ ,
\eeq
where $E_{\rm pk}$ is a reference energy chosen, for convenience, to be the peak of the Hawking distribution $d^2 N_\trg / dt dE_\trg$, \ie, the most common triggeron energy. The $\sigma_\ann(E_\trg) / \sigma_\ann(E_{\rm pk})$ function depends on the details of the DS model in question. We can rely on the fact that the cross section is invariant under boosts along the axis of collision to go to the CF, where $d \sigma_\ann / d\Omega \propto p_\med^{\rm [CF]} / \bl( s_M \, p_\pDM^{\rm [CF]} \br) \abs{\mathcal{M}}^2$; here we can use \Eqs{eq:Emstar}{eq:ann_boost} to find $p_\med^{\rm [CF]} = \sqrt{\bl( E_\med^{\rm [CF]} \br)^2 - m_\med^2}$ and $p_\pDM^{\rm [CF]} = \beta_\ann \gamma_\ann m_\pDM$. For an annihilation amplitude $\mathcal{M}$ independent of $E_\trg$, as is the case in the BaDM model, this is exactly the energy scaling of $\sigma_\ann$. In \Fig{fig:Lambda} we show the impact of an energy-dependent $\Lambda(E_\trg)$ on the differential photon flux relative to a constant $\Lambda = \Lambda_{\rm pk}$, assuming this $p_\med^{\rm [CF]} / \bl( s_M \, p_\pDM^{\rm [CF]} \br)$ energy scaling for both the BaDM and DTDM models. It is clear that the main effect is simply to suppress the amplitude of the signal at the high- and low-energy ends of the photon energy distribution. This is because the photons with the most extreme energies come from highly-boosted mediators, which in turn are only produced for sufficiently large triggeron energies. Since $\sigma_\ann(E_\trg)$ decreases with $E_\trg$, larger triggeron energies correspond to smaller $\Lambda(E_\trg)$ and therefore increasingly significant $\mathcal{J}$ suppressions as well. Triggeron energies both smaller and larger than $E_{\rm pk}$ are by definition less common, and thus have a smaller impact on the observed photon flux. Therefore, most of the effect of finite $\Lambda$ is captured by the typical $\Lambda_{\rm pk}$ constant value, with the energy-dependence of the cross section becoming important only at the edges of the photon distribution, away from its peak.

\section{Details of DTDM Model}\label{app:C}

In this appendix, we describe how the dipole interaction between the $\chi_1$ and the $\chi_2$ states of the DTDM model can be obtained. Let us consider a dark gauge symmetry $\SU{2}^\prime$ spontaneously broken by the VEV of a dark Higgs doublet $H^\prime$, as well as dark Dirac fermions $\chi = \bl( \chi_u , \, \chi_d \br)^T$ in a doublet of $\SU{2}^\prime$; and $\psi$, a singlet. The mass terms of the fermions are
\bea
    \mathcal{L}_{\rm DS} \subset \mathcal{L}_{\rm masses} = 
    && - M_\chi \anti{\chi} \chi - M_\psi \anti{\psi} \psi
    - y_{dL} \anti{\psi} H^{\prime \dagger} P_L \chi - y_{uL} \anti{\psi} \tilde{H}^{\prime \dagger} P_L \chi \nonumber\\
    && + \bl( L \leftrightarrow R \br)  \ + \ \mathrm{h.c.} \label{eq:mass_terms}
\eea
which, in all generality, are parity-breaking. In the $M_\psi \gg M_\chi, \, v^\prime$ limit, where $\langle H^\prime \rangle = (0, v^\prime / \sqrt{2})^T$ is the $\SU{2}'$-breaking VEV, we can integrate out $\psi$, which results in a mass-mixing matrix between $\chi_u$ and $\chi_d$:
\bea
    \mathcal{L}_{\rm \chi-mass} & = & \bl( \anti{\chi}_{uL}, \ \anti{\chi}_{dL} \br) \mathbf{M}
    \begin{pmatrix}
    \chi_{uR} \\
    \chi_{dR}
    \end{pmatrix} + \mathrm{h.c.} \ ,
    \nonumber\\
    \text{where} \quad \mathbf{M} & = & M_\chi \mathbf{1} - \delta m \ 
    \begin{pmatrix}
    y_{uL}^* y_{uR} & y_{uL}^* y_{dR} \\
    y_{dL}^* y_{uR} & y_{dL}^* y_{dR}
    \end{pmatrix} \\
    \text{and} \quad \delta m & \equiv & \frac{v^{\prime 2}}{2 M_\psi} \ . \label{eq:mass_mixings}
\eea
The mass matrix $\mathbf{M}$ can be diagonalized with chiral $\SU{2}'$ unitary rotations $\mathbf{R}$ and $\mathbf{L}$ in order to obtain the $\chi_m = \bl( \chi_2, \chi_1 \br)^T$ mass eigenstates: $\chi_R = \mathbf{R} \, \chi_{m R}$, and $\chi_L = \mathbf{L} \, \chi_{m L}$. In the simpler case of parity-preserving real Yukawas, $y_L = y_R = y \in \mathbb{R}$, we have $m_2 = M_\chi$, $\Delta m = (y_u^2 + y_d^2) \delta m$, and $\mathbf{L} = \mathbf{R} = \bl(  \bl( \cos\theta, -\sin\theta \br), \bl( \sin\theta, \cos\theta \br) \br)$ are orthogonal matrices with $\tan\theta = y_u/y_d$. If $\theta \ll 1$ then $\chi_1$ is mostly $\chi_d$, and $\chi_2$ mostly $\chi_u$. In the more general, parity-violating case, we still have $\Delta m \sim y^2 \delta m$ and $m_{1,2} \sim M_\chi$ for $y^2 \delta m \ll M_\chi$, with an $\mathcal{O}(1)$ mixing angle for generic Yukawas.

The dark Higgs VEV gives a mass $m_W' = g' v'$ to the $W^{a \prime}$ $\SU{2}'$ gauge bosons, where $g'$ is the dark gauge coupling. Furthermore, in the mass basis the $W'-\chi_m$ interactions are chiral and depend on $\mathbf{L}$ and $\mathbf{R}$, with the new kinetic term being
\beq
    \anti{\chi} i \cancel{D} \chi = \anti{\chi}_m \mathbf{L}^\dagger i \cancel{D} \mathbf{L} P_L \chi_m + \anti{\chi}_m \mathbf{R}^\dagger i \cancel{D} \mathbf{R} P_R \chi_m \ . \label{eq:new_kin}
\eeq
It is these $W'$ bosons that act as the triggerons. Indeed, the $W'$ from PBH Hawking radiation will trigger the annihilations $\chi_1 W^{a\prime} \to \chi_2 W^{b\prime}$, which readily convert the lighter $\chi_1$ DM particles into their heavier $\chi_2$ partners. Note that, for general $\mathbf{L}$ and $\mathbf{R}$, all three gauge bosons participate in the scattering, and thus all three can be considered to be triggerons. The coupling of these triggerons to the $\chi_m$ mass eigenstates involves the matrices $\mathbf{L}^\dagger \tau^a \mathbf{L}$ and $\mathbf{R}^\dagger \tau^a \mathbf{R}$; this is what is meant by $V$ in \Eq{eq:sigma_chiW}. The benchmark values of $m_{1,2} \sim 10~\MeV$, $m_{W'} \sim 1~\MeV$, $\Delta m \sim 0.1~\MeV$, and $g' \sim \mathcal{O}(\text{few})$ used in Subsec.~\ref{subsec:dtdm} can be readily obtained with $M_\chi \sim 10~\MeV$, $v' \sim 1~\MeV$, and $M_\psi \sim 100~\MeV \, (y/3)^2$.

\begin{figure}\centering
\includegraphics[width=0.6\linewidth]{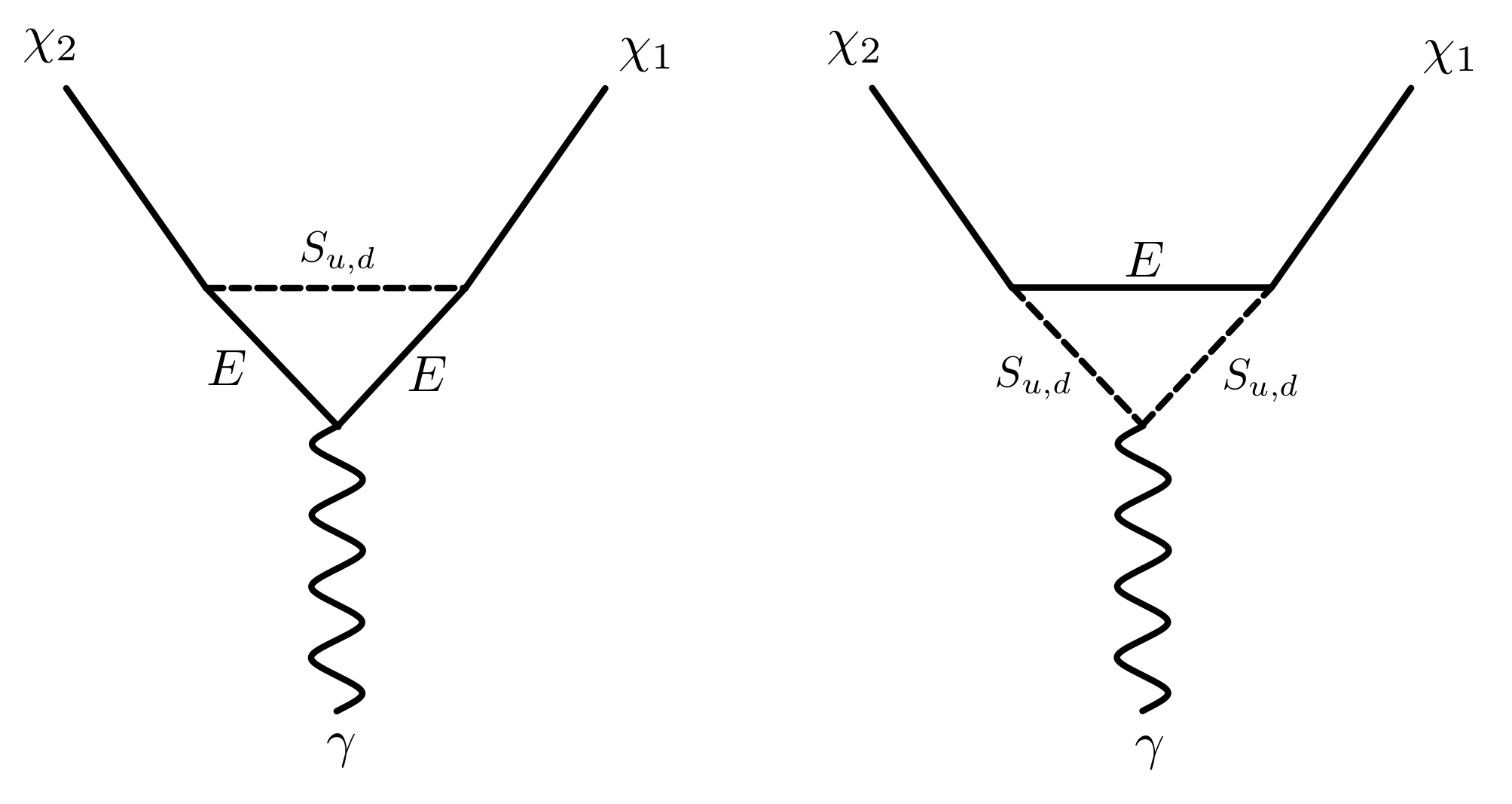}
\caption{Loop diagrams generating the inelastic dipole coupling to the SM photon of \Eq{eq:dipole}. Note that the external fermions $\chi_{1,2}$ are the mass eigenstates $\chi_m$, which are a linear combination of the up and down components of the original $\chi$ $\SU{2}'$ doublet. This means that the $\chi_m$ vertices in these diagrams have factors of $\mathbf{L}$ and $\mathbf{R}$, as per \Eq{eq:dipole_terms}. The heavy scalars $S_{u,d}$ are charged under $\Ua_{\rm QED}$ and are also part of a $\SU{2}'$ doublet, while the heavy fermion $E$ is only charged under $\Ua_{\rm QED}$.}
\label{fig:dipole_diagrams}
\end{figure}

The dipole term $d \anti{\chi}_2 \sigma^{\mu\nu} \chi_1 F_{\mu\nu}$ can be generated at 1-loop level in the presence of interactions involving $\chi$ and, for example, a $\SU{2}^\prime$ scalar doublet $S$ with $\Ua_{\rm QED}$ charge $q'$, as well as a vector-like fermion $E$ of charge $-q'$.\footnote{Above the EW scale, we should think of these fields as being charged under weak hypercharge $\Ua_Y$ instead.} Thus, we add to $\mathcal{L}_{\rm DS}$ the following terms
\bea
    \mathcal{L}_{\rm dipole} & = & - \lambda_L \anti{E} S^\dagger P_L \chi + \bl( L \leftrightarrow R \br) + \mathrm{h.c.} \nonumber\\
    & = & - \lambda_L \anti{E} S^\dagger \mathbf{L} P_L \chi_m + \bl( L \leftrightarrow R \br) + \mathrm{h.c.} \ . \label{eq:dipole_terms}
\eea
These give rise to the inelastic dipole between $\chi_1$ and $\chi_2$ through the Feynman diagrams shown in \Fig{fig:dipole_diagrams}:
\bea
    d \!\! & \approx & \!\! \frac{\lambda_R^* \lambda_L q' e}{16 \pi^2} \bl( \bl[ \mathbf{R}^\dagger \mathbf{L} \br]_{du} + \bl[ \mathbf{L}^\dagger \mathbf{R} \br]_{du} \br) \frac{M_E}{M_S^2} \nonumber\\
    \!\! & \approx & \!\! 10^{-8}~\GeV^{-1} \!\! \bl( \frac{\lambda_R^* \lambda_L q' U}{0.5} \br) \!\!\! \bl( \frac{M_E}{1~\TeV} \br) \!\!\! \bl( \frac{10~\TeV}{M_S} \br)^2 \ , \label{eq:dipole}
\eea
where $M_S$ and $M_E$ denote the masses of $S$ and $E$ respectively, and $U \equiv \bl[ \mathbf{R}^\dagger \mathbf{L} \br]_{du} + \bl[ \mathbf{L}^\dagger \mathbf{R} \br]_{du}$.

It is clear from this that in the parity-preserving case $\mathbf{L} = \mathbf{R} \Rightarrow U = 2 \cdot \bl[ \mathbf{1} \br]_{du} = 0$, and thus the dipole interaction vanishes. This can be understood in terms of a residual global $\SU{2}'$ symmetry after $H'$ gets a VEV, which allows us to redefine $S \to \mathbf{L}^\dagger S$ when rotating $\chi$ into $\chi_m$ in \Eq{eq:dipole_terms}. As a result, we have a new $S = (S_u, S_d)^T$ whose components couple to $(\chi_2, \chi_1)$ without any mixing, which means the $S$ in the loops of \Fig{fig:dipole_diagrams} can never take $\chi_2$ into $\chi_1$. The more general parity-breaking lagrangian of \Eq{eq:mass_terms} explicitely breaks such a global symmetry. Another way to couple {\it each} of $S_{u,d}$ to {\it both} $\chi_{1,2}$ while preserving parity is to split the scalar masses. This can be done with the terms $\kappa \vert H^{\prime \dagger} S \vert^2 + \tilde{\kappa} \vert \tilde{H}^{\prime \dagger} S \vert^2 + \mathrm{h.c.}$, which generate a mass splitting $\Delta M_S^2 \sim \kappa v^{\prime2}$. Such a splitting means that we cannot redefine $S$ using $\mathbf{L}$, and thus the mixing matrix $\mathbf{L}$ remains in \Eq{eq:dipole_terms} after going to the $\chi_m$ mass eigenstates. Since now $S_{u,d}$ would have different masses and thus different propagators, their contributions to the loops in \Fig{fig:dipole_diagrams} do not cancel exactly, and a dipole is generated, equal to that in \Eq{eq:dipole} multiplied by a suppression $( \Delta M_S^2 / M_S^2 )^2$. Because we are considering PBHs with temperatures $T_\PBH \sim \mathcal{O}(1\text{--}10~\MeV)$ and we need $m_W' = g' v' < T_\PBH$ and $g' \sim \mathcal{O}(1)$, our benchmarks give an enormous suppression of $( \Delta M_S^2 / M_S^2 )^2 \lesssim 10^{-28}$, which means that $\chi_2$ would not decay on cosmological timescales. Therefore, we simply assume that parity in the dark sector, as in the visible sector, is violated.

\bibliographystyle{utphys}
\bibliography{reference.bib}

\end{document}